\documentclass{ptephy_v1}
\preprintnumber{XXXX-XXXX} 

\begin{document}

\title{First neutrino event detection with nuclear emulsion at J-PARC neutrino beamline}


\author{\name{T. Fukuda}{1,*}, \name{S. Aoki}{2}, \name{S. Cao}{3}, \name{N. Chikuma}{4}, \name{Y. Fukuzawa}{1}, \name{M. Gonin}{5}, \newline \name{T. Hayashino}{6}, \name{Y. Hayato}{7}, \name{A. Hiramoto}{6}, \name{F. Hosomi}{4}, \name{K. Ishiguro}{1}, \name{S. Iori}{8}, \newline \name{T. Inoh}{8}, \name{H. Kawahara}{1}, \name{H. Kim}{9}, \name{N. Kitagawa}{1}, \name{T. Koga}{4}, \name{R. Komatani}{1}, \newline \name{M. Komatsu}{1}, \name{A. Matsushita}{1}, \name{S. Mikado}{10}, \name{A. Minamino}{11}, \name{H. Mizusawa}{8}, \newline \name{K. Morishima}{1}, \name{T. Matsuo}{8}, \name{T. Matsumoto}{8}, \name{Y. Morimoto}{8}, \name{M. Morishita}{1}, \newline \name{K. Nakamura}{6}, \name{M. Nakamura}{1}, \name{Y. Nakamura}{1}, \name{N. Naganawa}{1}, \name{T. Nakano}{1}, \newline \name{T. Nakaya}{6}, \name{Y. Nakatsuka}{1}, \name{A. Nishio}{1}, \name{S. Ogawa}{8}, \name{H. Oshima}{8}, \name{B. Quilain}{6}, \newline \name{H. Rokujo}{1}, \name{O. Sato}{1}, \name{Y. Seiya}{9}, \name{H. Shibuya}{8}, \name{T. Shiraishi}{1}, \name{Y. Suzuki}{1}, \name{S. Tada}{1}, \newline \name{S. Takahashi}{2}, \name{K. Yamada}{2}, \name{M. Yoshimoto}{1} and \name{M. Yokoyama}{4}}

\address{\affil{1}$^{1}${Nagoya University, Nagoya 464-8601, Japan} \\
\affil{2}$^{2}${Kobe University, Kobe 657-8501, Japan} \\
\affil{3}$^{3}${High Energy Accelerator Research Organization (KEK), Tsukuba 305-0801, Japan} \\
\affil{4}$^{4}${University of Tokyo, Tokyo 113-0033, Japan} \\
\affil{5}$^{5}${Ecole Polytechnique, IN2P3-CNRS, Laboratoire Leprince-Ringuet, Palaiseau, France} \\
\affil{6}$^{6}${Kyoto University, Kyoto 606-8502, Japan} \\
\affil{7}$^{7}${University of Tokyo, ICRR, Kamioka 506-1205, Japan} \\
\affil{8}$^{8}${Toho University, Funabashi 274-8510, Japan} \\
\affil{9}$^{9}${Osaka City University, Osaka 558-8585, Japan} \\
\affil{10}$^{10}${Nihon University, Narashino 274-8501, Japan} \\
\affil{11}$^{11}${Yokohama National University, Yokohama 240-0067, Japan} \\\email{tfukuda@flab.phys.nagoya-u.ac.jp}}

\begin{abstract}%
Precise neutrino--nucleus interaction measurements in the sub-multi GeV region are important to reduce the systematic uncertainty in future neutrino oscillation experiments. Furthermore, the excess of ${\nu_e}$ interactions, as a  possible interpretation of the existence of a sterile neutrino has been observed in such an energy region. 
The nuclear emulsion technique can measure all the final state particles with low energy threshold for a variety of targets (Fe, C, H${_2}$O, and so on). Its sub-$\mu$m position resolution allows measurements of the ${\nu_e}$ cross-section with good electron/gamma separation capability.
We started a new experiment at J-PARC to study sub-multi GeV neutrino interactions by introducing the nuclear emulsion technique. The J-PARC T60 experiment has been implemented as a first step of such a project. Systematic neutrino event analysis with full scanning data in the nuclear emulsion detector was performed for the first time. The first neutrino event detection and its analysis is described in this paper.
\end{abstract}

\subjectindex{xxxx, xxx}

\maketitle

\section{Introduction}

Currently, many experimental neutrino oscillation projects are running or planned to search for the CP violation, and/or to probe the neutrino mass hierarchy \cite{t2k,hk,nova,dune,juno,reno50,ino,pingu,km3net}. Precise cross-section measurements of neutrino-nucleus interactions in the sub-multi GeV region are important to reduce the systematic uncertainty, specially for neutrino mixing angle $\theta_{23}$, in these experiments. Furthermore, one possible explanation of the MiniBooNE anomaly \cite{miniboone} in this energy region is a result of the existence of so-called sterile neutrino indicated by LSND \cite{lsnd}. Therefore that is also important to verify the MiniBooNE anomaly.

Nuclear emulsion is a three-dimensional solid tracking detector with sub-micron positional resolution. Thanks to its high spatial resolution, the nuclear emulsion technology contributed to the discovery of the pion \cite{pion}, of the charmed particle in cosmic rays \cite{charm}, the direct observation of ${\nu_{\tau}}$ \cite{donut}, the discovery of ${\nu_{\tau}}$ appearance in a ${\nu_{\mu}}$ beam \cite{opera} and so on. This high spatial resolution allows measurements of all the final state particles with low-energy threshold. Nuclear emulsion has 4$\pi$ track detection capability \cite{LA}. This is also very useful to investigate the event topology in low energy neutrino interactions. Furthermore, good $e/\gamma$ separation capability allows measurement of ${\nu_e}$ CC interactions with a strong suppression of the background from ${\nu_{\mu}}$ NC interactions with $\pi^0$ production. There is much flexibility for target material selection because the detector, a so-called Emulsion Cloud Chamber (ECC), is constructed as a sandwich structure of thin nuclear emulsion films and the target materials.

The J-PARC T60 experiment \cite{t60} was proposed to study the feasibility and check the detector performance of nuclear emulsion at the J-PARC neutrino beamline. The first results of beam exposure in 2015 will be presented in this paper.

Accelerator-based neutrino experiments with nuclear emulsion were established from the end of 1970s \cite{exxx,e531,chorus}. The experimental concept is that of a hybrid analysis with the emulsion and the electronic detectors, so-called "Emulsion-Counter Hybrid Experiment". Only a small area of emulsions that is predicted by the electronic detectors has been analyzed, or only a small number of tracks that are predicted have been traced back because of the limitation of the track readout speed in the emulsions. The high-speed scanning system of the emulsion has been developed \cite{ts,hts} and the track readout of the entire area in the emulsion has recently become available. This allows one to analyze the emulsions independently of the prediction from the electronic detectors. In this paper, the first systematic neutrino event analysis with scanning all data in the nuclear emulsion will be also described.

\section{Neutrino Detector}

\subsection{Nuclear Emulsion Films}
\begin{figure}[ht]
\begin{center}
\includegraphics[clip, width=14.0cm]{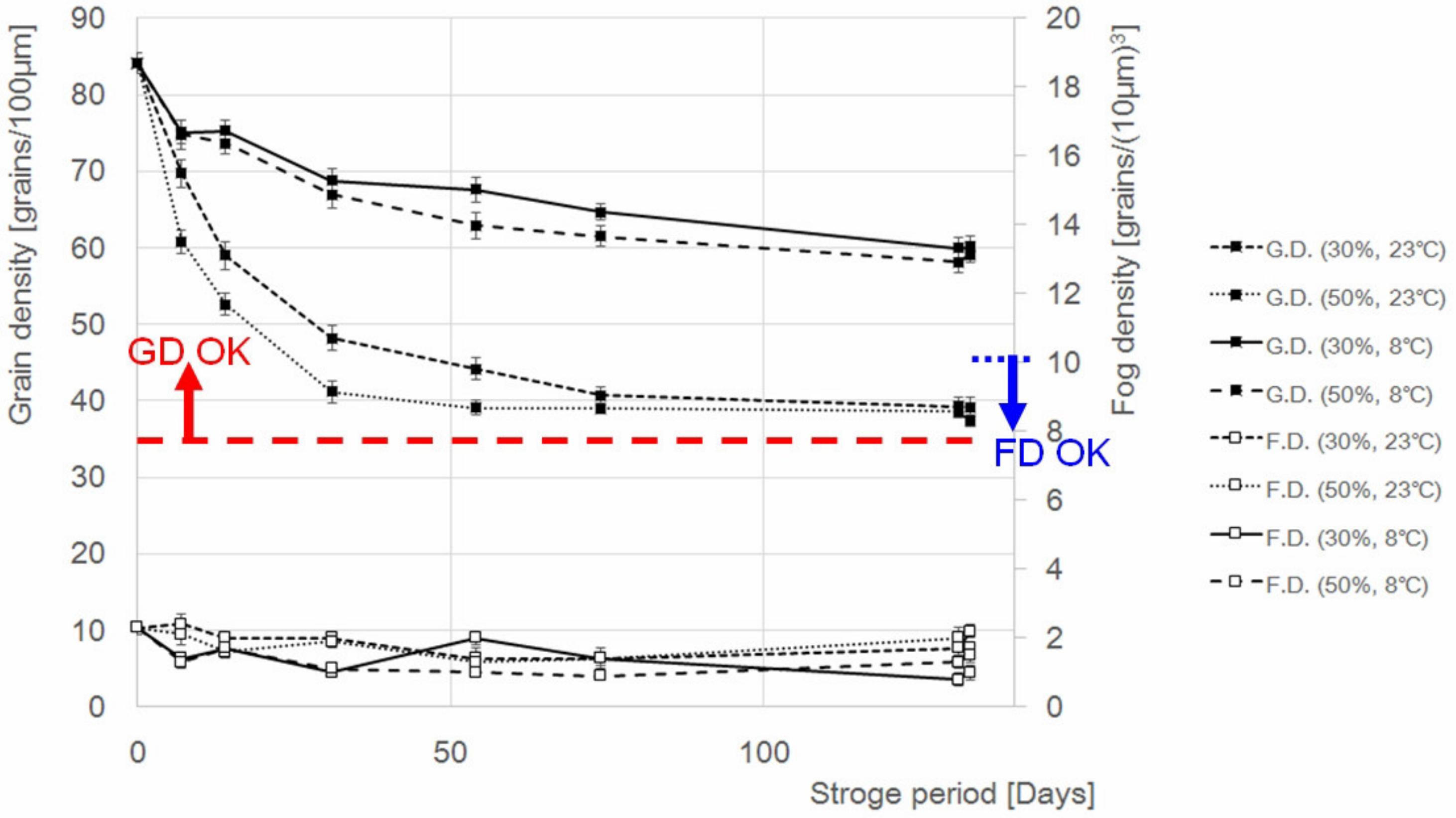}
\caption{Initial and long-term performance of emulsion gels. The GD and FD is kept at a safe levels ( GD $\geq$ 35, FD $\leq$ 10 ) for 133 days at less than 23~$^\circ$C. The fading effect is suppressed in low-temperature and low-humidity conditions.}
\label{emulsion}
\end{center}
\end{figure}

Nuclear emulsion gel was produced at the facility of Nagoya University. In this emulsion gel production, high-sensitivity emulsion gel, containing 55\% AgBr crystal by volume ratio, was produced \cite{highGD}. Two emulsion layers, each 50 $\mu$m thick, were formed on both faces of a 180-$\mu$m thick polystyrene base. The size of sheets was 100 mm $\times$ 125 mm. The long-term variation of the emulsion sensitivity and background noise was investigated using cut pieces of emulsion sheets. The emulsion sensitivity was measured as grain density (GD), the number of grains per 100 $\mu$m along a minimum ionizing particle (MIP) trajectory. The noise of the emulsion was measured as fog density (FD), the number of randomly distributed grains per 1000 $\mu m^3$ in the emulsion. If an emulsion is left undeveloped for a long time after track recording, the silver atoms tend to oxidize into silver ions, owing to the presence of water and oxygen in the material. Some of the latent image cores are destroyed and fewer grains are produced in the development process. This effect is called fading. Variations of GD and FD over 133 days are shown in Fig.\ref{emulsion} under four different temperature and humidity conditions. We found that GD and FD remain at a safe levels, even after 133 days at 23~$^\circ$C.

\subsection{Emulsion Cloud Chamber (ECC)}
\begin{figure}[ht]
\begin{center}
\includegraphics[clip, width=8.0cm]{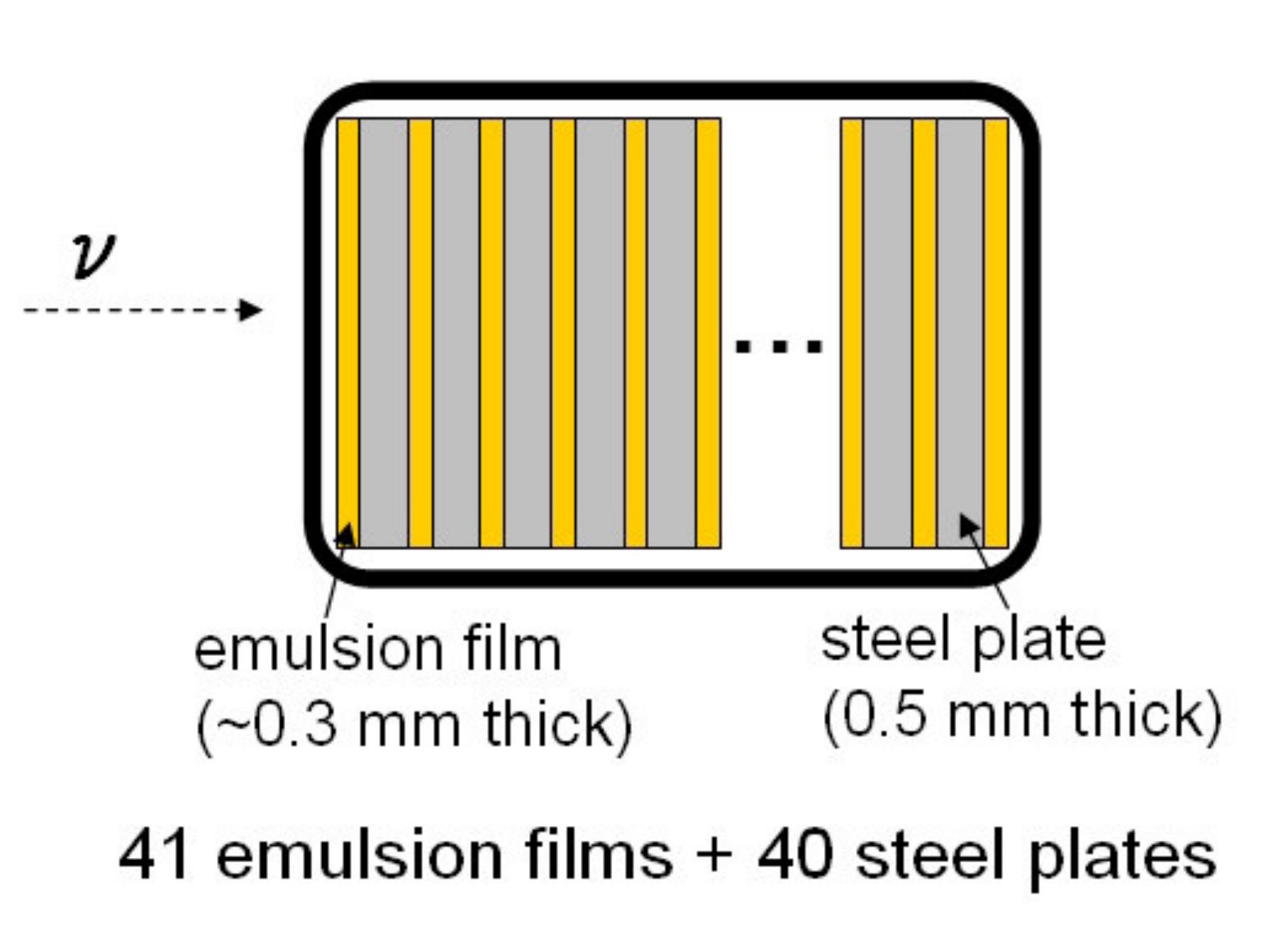}
\caption{The Emulsion Cloud Chamber (ECC)}
\label{ECC}
\end{center}
\end{figure}

The ECC is constituted by alternatively stacking 41 emulsion films and 40 stainless steel plates (SUS304, 500 $\mu$m thickness) as shown in Fig.\ref{ECC}. These are inserted in a paper box and then vacuum-packed with a light-shield aluminum laminated bag at 40\% R.H. condition. The total target mass is approximately 2.2 kg (1.2 kg inside the fiducial volume).

\newpage

\subsection{Emulsion Shifter}
\begin{figure}[ht]
\begin{center}
\includegraphics[clip, width=9.0cm]{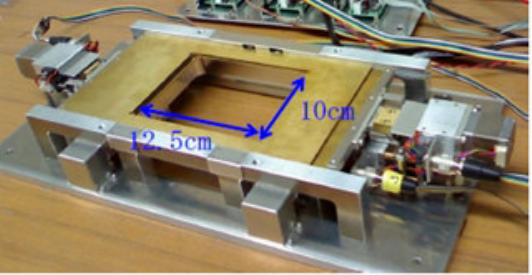}
\caption{Emulsion shifter}
\label{shifter}
\end{center}
\end{figure}

Precise tracking detectors with timing information, e.g. the silicon microstrip detector or the scintillating fiber detector, have been applied to connect tracks between the main emulsion detector and the electronic detectors, which have muon-identification capability in the conventional Emulsion-Counter Hybrid Experiment. In the present work, an emulsion device with timing information played this role. The emulsion detector intrinsically has no timing information. Therefore, the timing information is produced from the emulsion tracks by track coincidence between the films on the a mechanically controlled moving stage \cite{shifter01,shifter02}, a so-called the emulsion shifter. 
An emulsion shifter (Fig.\ref{shifter}) was reused from a balloon experiment with nuclear emulsion \cite{shifter}. This equipment gives a time-stamp to the emulsion track by using a clock-based multi-stage emulsion shifter technique. Three stages play the roles of the hour, minute and second hands of a watch, respectively. The emulsion tracks that are coincident on the all three stages obtained the time-stamp corresponding to the each stage position. The timing resolution of this emulsion shifter is approximately 7.9 seconds, which was investigated using tracks passing through the emulsion shifter and electronic detector located at the back of the emulsion detectors. The details for this equipment and the analysis are described in \cite{shifter,t60shifter}. 

\subsection{Interactive Neutrino GRID (INGRID)}
\begin{figure}[hbt]
\begin{center}
\includegraphics[clip, width=11.0cm]{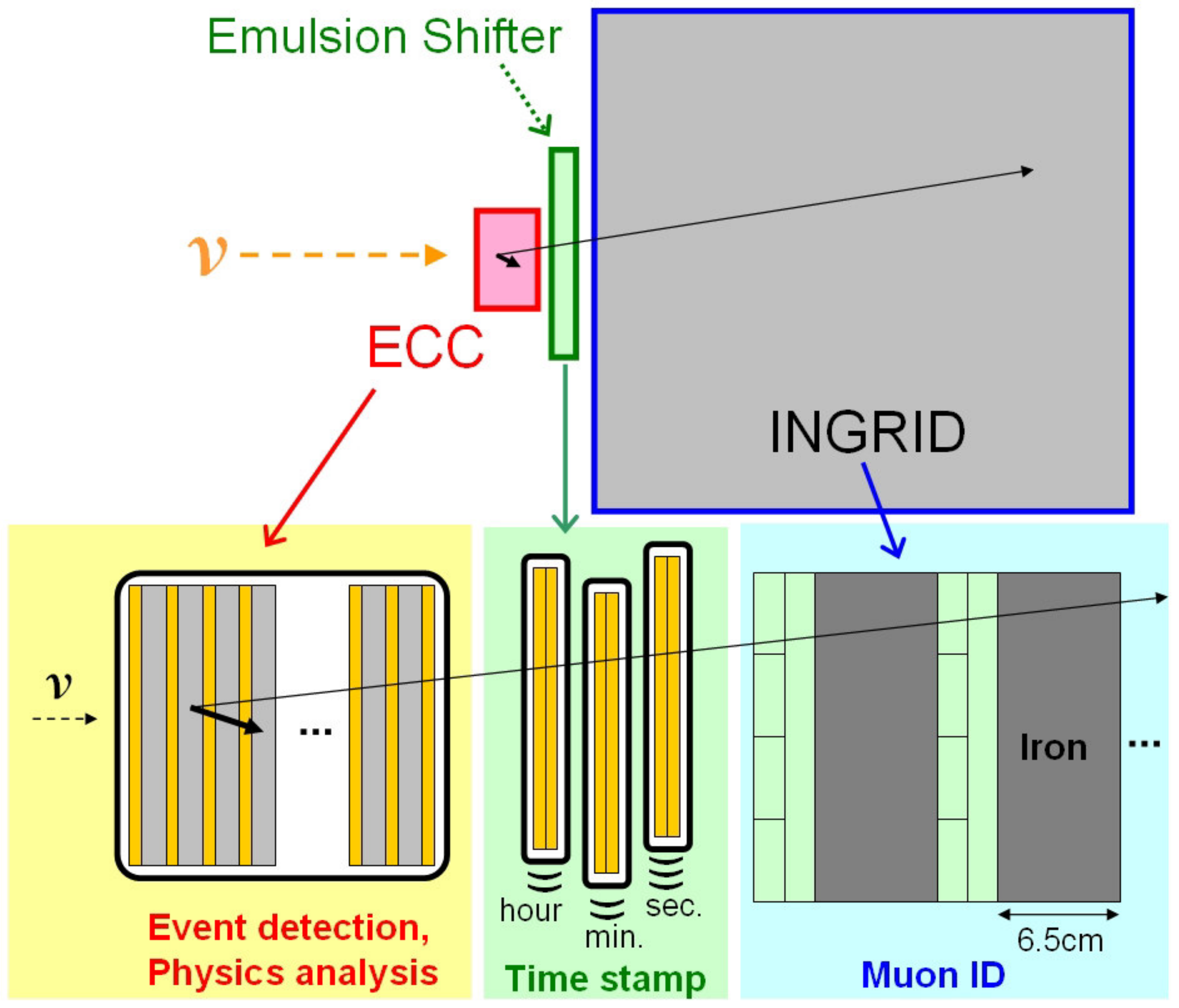}
\caption{Detector concept}
\label{concept}
\end{center}
\end{figure}

INGRID is one of the near detectors for the T2K experiment \cite{ingrid}. The tracking device is a 1-cm-thickness plastic scintillator (5 cm width) read out with wave length-shifting fibers. The constitution of the detector is a sandwich structure of the tracking planes and 6.5-cm-thickness iron blocks. The detector was assembled from a total of nine iron layers (58.5 cm) and 11 vertical and horizontal tracking planes. The size of an module is 1.2 m $\times$ 1.2 m $\times$ 0.9 m. INGRID consists of 16 identical modules arranged in horizontal and vertical arrays around the beam center. The original purpose of INGRID is to measure the neutrino beam direction and verify the number of neutrino interactions per protons on target. In the J-PARC T60 experiment, INGRID is utilized to identify the muons from neutrino--nucleus interactions in the ECC. As shown in Fig.\ref{concept}, the emulsion shifter is placed between the ECC and INGRID. The particles that cross the emulsion shifter are supplied time-stamps. The tracks are then matched between the emulsion detector and INGRID using timing information.

\subsection{Installation at J-PARC}
\begin{figure}[hbt]
\begin{center}
\includegraphics[clip, width=13.0cm]{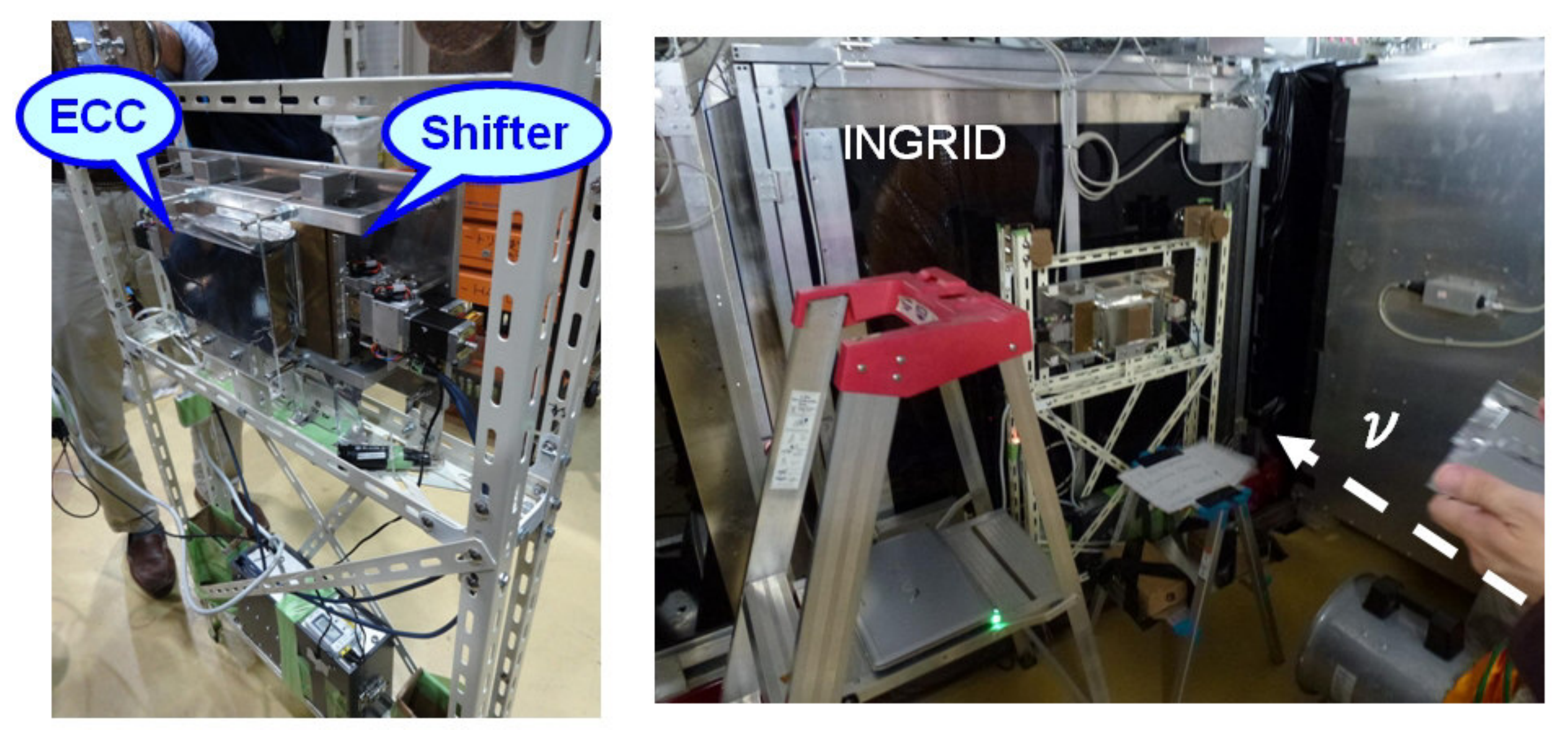}
\caption{Detector installation at J-PARC neutrino beamline}
\label{installation}
\end{center}
\end{figure}

The emulsion detectors and the assembling frame were prepared at Toho University. These detectors were transported to J-PARC before the installation. The emulsion detectors were installed in front of an INGRID module adjacent to the on-axis module of the Neutrino Monitor Building in J-PARC on 14th Jan. 2015, as shown in Fig.\ref{installation}. The operation of the emulsion shifter has been monitored by verifying the stage reproducibility, which was sent to Kobe University with a monitoring laptop PC.

\section{Neutrino Beam Exposure}

\begin{figure}[ht]
\begin{center}
\includegraphics[clip, width=13.0cm]{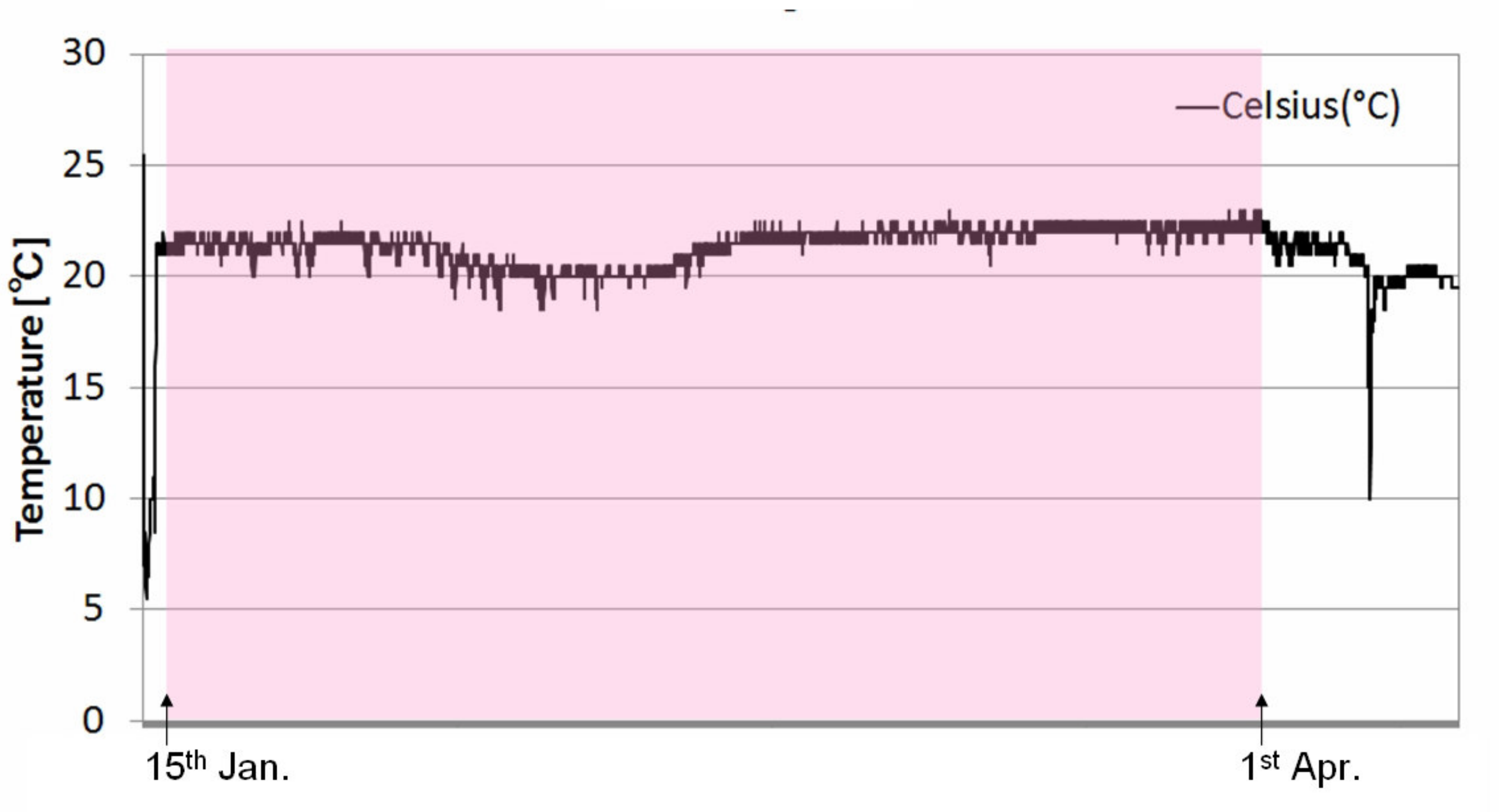}
\caption{Temperature at the experimental site}
\label{temp}
\end{center}
\end{figure}

The neutrino beam exposure started in 15th Jan. 2015. Unfortunately, beam exposure was stopped on 16th Jan., then re-started from 25th Feb. and finished on 1st Apr. The beam exposure was performed with the anti-neutrino beam mode after re-start of beam exposure. The accumulated number of protons on target is approximately 13.8 ${\times}$ 10$^{19}$. The ratio of neutrino and anti-neutrino interactions is $\sim$ 1:3 because of the contamination of neutrinos in anti-neutrino beam. The total number of neutrino or anti-neutrino events is estimated as approximately 44 events (Fiducial Volume: 24 events, including two events at the region of the emulsion layer). After beam exposure, the stages of the emulsion shifter were moved to a reference position and reference cosmic-ray tracks were accumulated for one week. The emulsion detectors were uninstalled on 8th Apr. and all films were immediately developed at Toho University. The temperature at the experimental site was approximately 22 ${}^\circ$C during the exposure, as shown in Fig.\ref{temp}. Therefore, the tracks that are recorded in emulsion films were kept in good condition, as shown in Fig.\ref{emulsion}. The operation of the emulsion shifter was well performed \cite{t60shifter}.

\clearpage

\section{Event Analysis}

\subsection{Scanning}
\begin{figure}[ht]
\begin{center}
\includegraphics[clip, width=10.0cm]{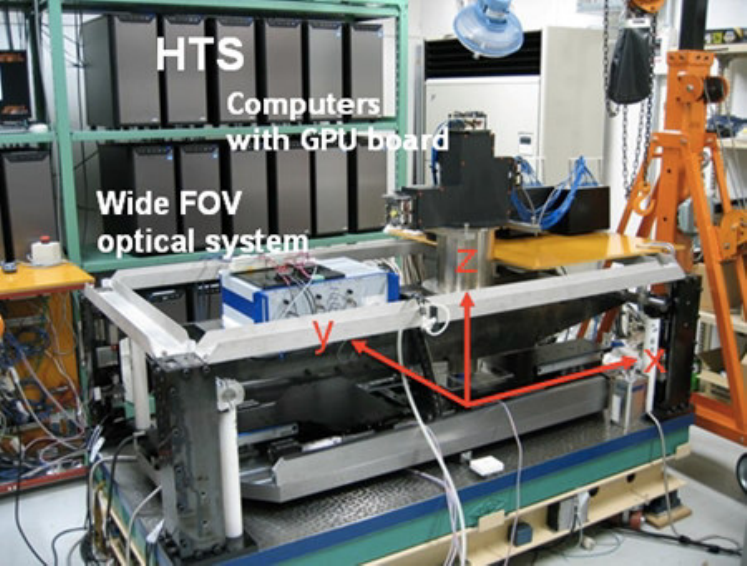}
\caption{Hyper Track Selector (HTS)}
\label{HTS}
\end{center}
\end{figure}

The entire areas of all the emulsion films for the experiment ($\sim$1.2 m$^2$ ; for two emulsion layers of all the emulsion films, including six films for the emulsion shifter) were scanned using the high-speed automated microscope system, Hyper Track Selector (HTS) \cite{hts} (Fig.\ref{HTS}). This system has been developed in Nagoya University to quickly analyze particle tracks in emulsion films and allows to readout tracks at a scanning speed of $\sim$5,000 cm$^2$/h. HTS takes 16 tomographic images in an emulsion layer and recognizes a series of grains on a straight line as a track, by taking a coincidence among these images. In this time, the slope acceptance of HTS is set at $|$tan$\theta$$|$ $<$ 1.7 ($|\theta| \lesssim$ 60 degree), where $\theta$ is the track angle with respect to the perpendicular of the emulsion film (the $z$-axis).

\subsection{Analysis in ECC}

\subsubsection{Track Reconstruction}
\begin{figure}[ht]
\begin{center}
\includegraphics[clip, width=15.0cm]{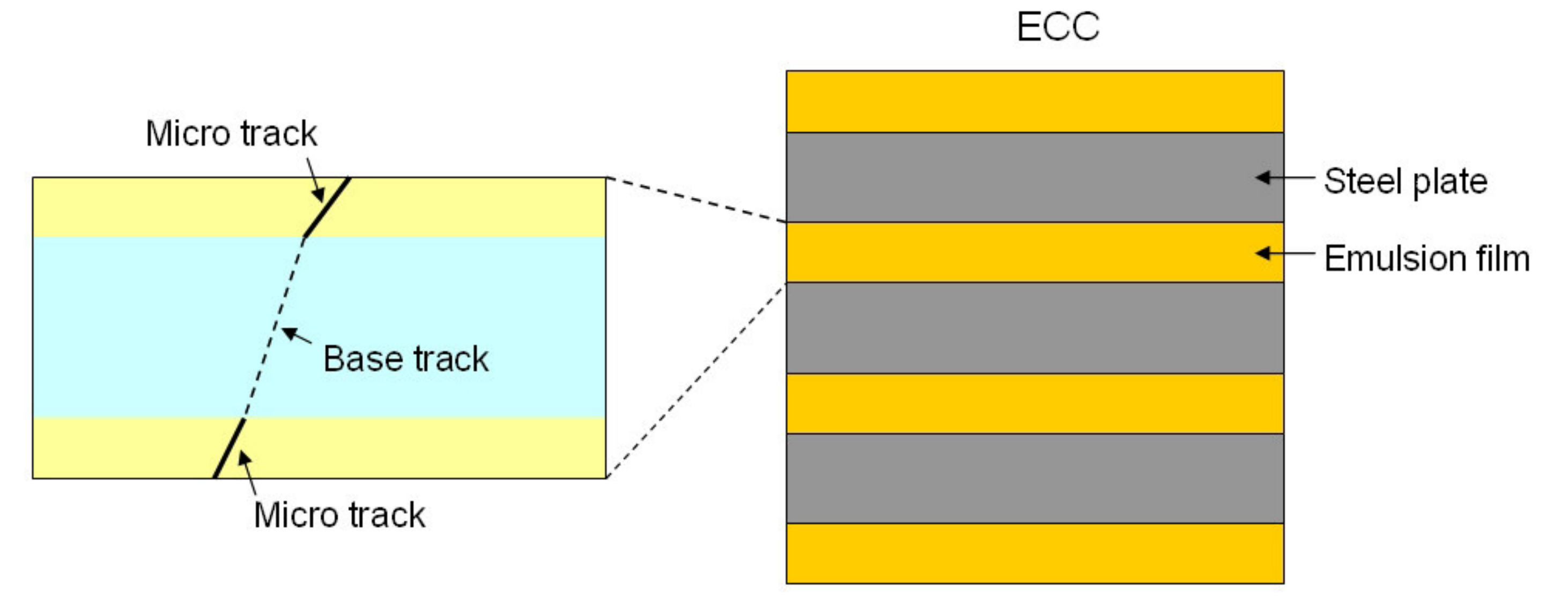}
\caption{Schematic view of the track reconstruction in an emulsion film.}
\label{trackdata}
\end{center}
\end{figure}

\begin{figure}[ht]
\begin{center}
\includegraphics[clip, width=15.0cm]{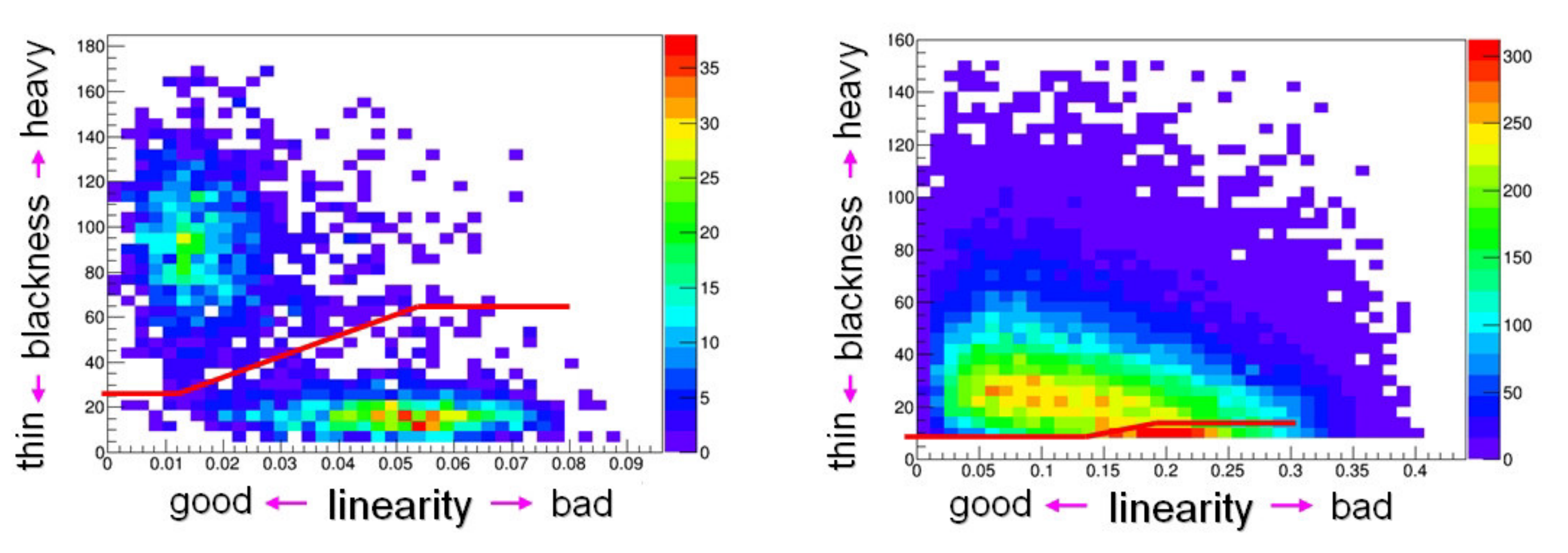}
\caption{Track linearity and blackness (left: $|$tan$\theta$$|$ $<$ 0.1, right: $|$tan$\theta$$|$ $>$ 1.4). The sum of VPH of micro tracks, which is a reconstructed base track is shown on the vertical axis. The square-root of the sum of squares of the angle difference between the base track and micro tracks is shown on the  horizontal axis. The blackness and linearity of fake tracks tend to be thin and bad, respectively. The red line shows the discrimination line and is set for each 100 mrad of track angle.}
\label{ranking}
\end{center}
\end{figure}

First, track segments, so-called ``micro tracks", were detected on each layer of the emulsion films (Fig.\ref{trackdata}). The positions ($x$, $y$), slopes (tan$\theta_x$, tan$\theta_y$), number of hits in an emulsion layer, Pulse Height (PH), and number of pixels constructed track data, Volume Pulse Height (VPH) of the micro tracks, were measured by HTS. The number of grains belonging to the track (blackness) is well correlated with the PH and the VPH. As shown in Fig.\ref{trackdata}, the tracks connecting the positions of micro tracks on both side of the plastic base are called ``base tracks". Sub-MeV or few MeV electrons from the environment are recorded in the emulsion because nuclear emulsion records all tracks from immediately after production until development with no dead time. Therefore, the automated scanning system recognizes ``fake" tracks made of low-energy electrons and/or random noise grains (fog). The chance coincidence of these fake tracks become the background of base tracks. However, such background base tracks can be rejected by evaluating their blackness (VPH) and linearity (the angle difference between base track and micro tracks) \cite{TRank}. Fig.\ref{ranking} shows the blackness vs. the linearity of base tracks for a small-angle region ($|$tan$\theta$$|$ $<$ 0.1 : left figure) and large-angle region ($|$tan$\theta$$|$ $>$ 1.4 : right figure). The VPH and angular accuracy of base tracks depend on their angle. The smaller the track angle, the better the signal-to-noise ratio (S/N) of the base tracks. So the discrimination line is set for each track angle as shown by the red line in Fig.\ref{ranking}. The tracks under the red line are rejected as the fake track.

\begin{figure}[h]
\begin{center}
\includegraphics[clip, width=12.0cm]{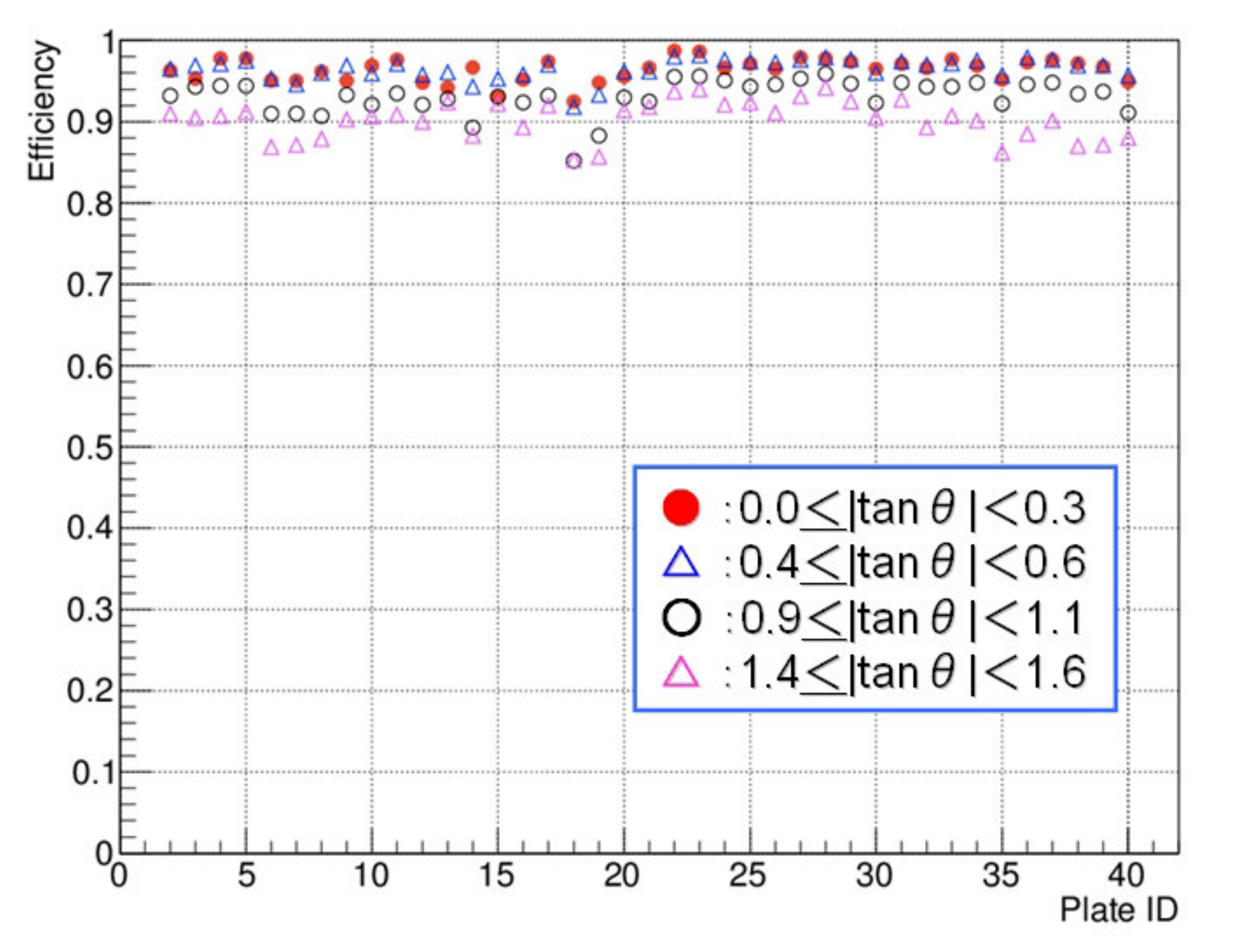}
\caption{Track efficiency after reconstruction. When the track efficiency is estimated at one plate, having base tracks at upstream and downstream plates is required to select penetrating tracks in the target plate.}
\label{eff}
\end{center}
\end{figure}

The positions ($x$, $y$) and slopes (tan$\theta_x$, tan$\theta_y$) of the base tracks were measured with respect to the microscope's coordinate system. After connecting the base tracks among the emulsion films, the rotation, slant, parallel translation, and distance between every pair of adjacent emulsion films are adjusted, so that the differences in the track positions and slopes between the two films are minimized after the reconstruction procedure. These connection elements of base tracks are called ``linklets". All base tracks in all films are reconstructed by tracking linklets, as the chain.  Finally, the track density in a film is $\sim$4,000 tracks / cm$^2$. The track efficiency after reconstruction was evaluated at each measured angle for all plates, as shown in Fig.\ref{eff}. In the small-angle region, the track efficiency is approximately 96 \%. However, in the large-angle region, the efficiency loss caused is found.

\vspace*{\stretch{1}}

\vspace{\stretch{1}}

\clearpage

\subsubsection{Proton Identification}
The momentum of charged particles can be estimated in the ECC by measuring their Multiple Coulomb Scattering (MCS) \cite{mcs}. When a particle of charge ${z}$, momentum ${p}$, velocity ${\beta c}$ traverses a material of depth ${x}$ and radiation length ${X_0}$, the deviation of the distribution of scattering angle is given by Eq.1.

\begin{equation}
\theta_0 = \frac{13.6MeV}{\beta cp}z\sqrt{\frac{x}{X_0}}(1+0.038 ln(\frac{x}{X_0}))
\label{mcs}
\end{equation}

\begin{figure}[ht]
\begin{center}
\includegraphics[clip, width=11.0cm]{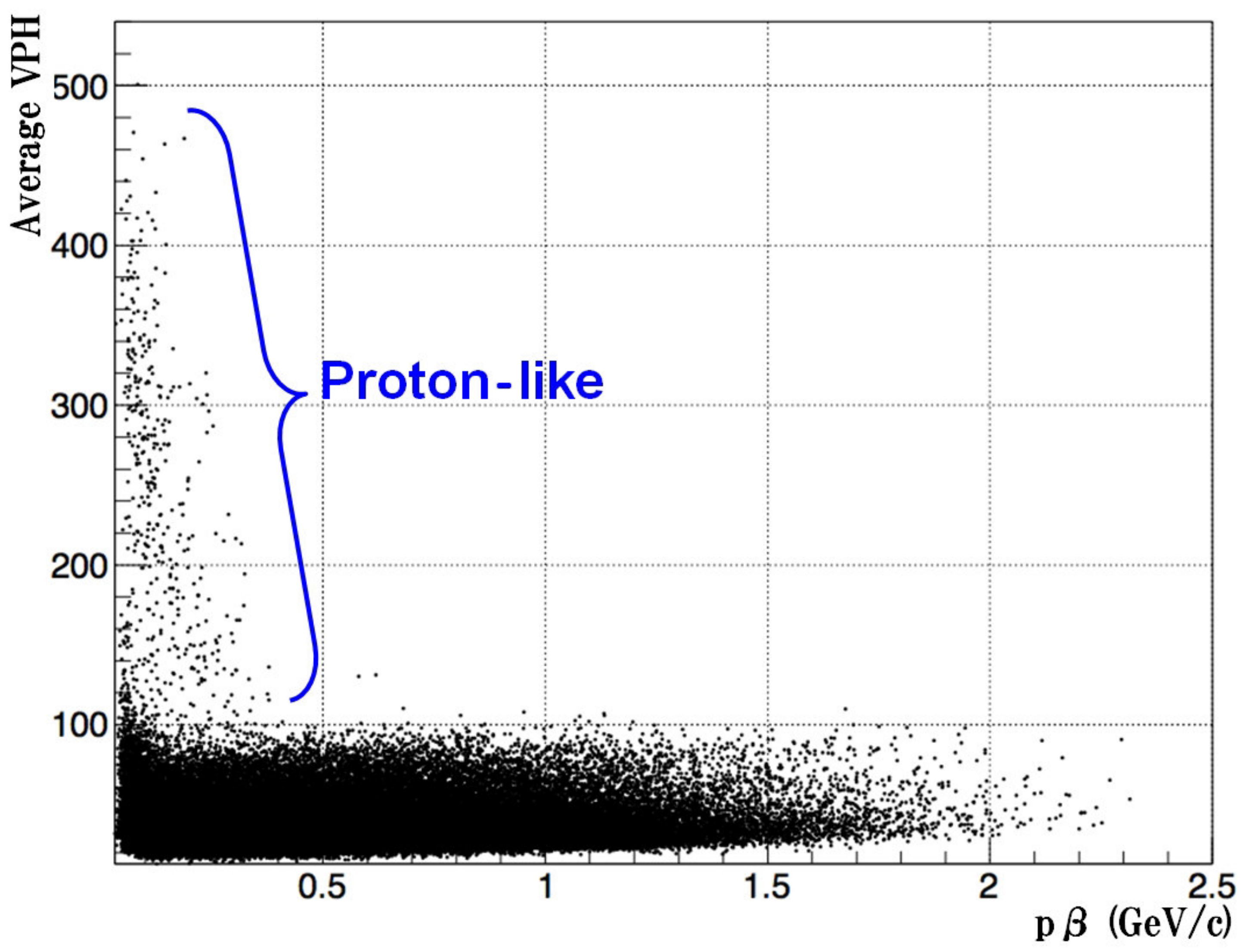}
\caption{Proton-like tracks are seen in the low $p\beta$ and high average VPH region.}
\label{pid}
\end{center}
\end{figure}

The GD is nearly proportional to its energy loss d$E$/d$x$ in nuclear emulsion \cite{gd}. The average number of hit layers (PH) or the average number of pixels associated to the reconstructed track (VPH) in scanned track data corresponds to the GD of tracks \cite{ppi}. In Fig.\ref{pid}, the measured momentum and the average VPH of each track that is connected to more than 10 segments is shown. The proton-like tracks are found in the low-momentum and high-d$E$/d$x$ region.

\subsubsection{Track Selection}

The reconstructed tracks are selected to search for neutrino event vertices as shown the following steps.

\begin{figure}[ht]
\begin{center}
\includegraphics[clip, width=13.0cm]{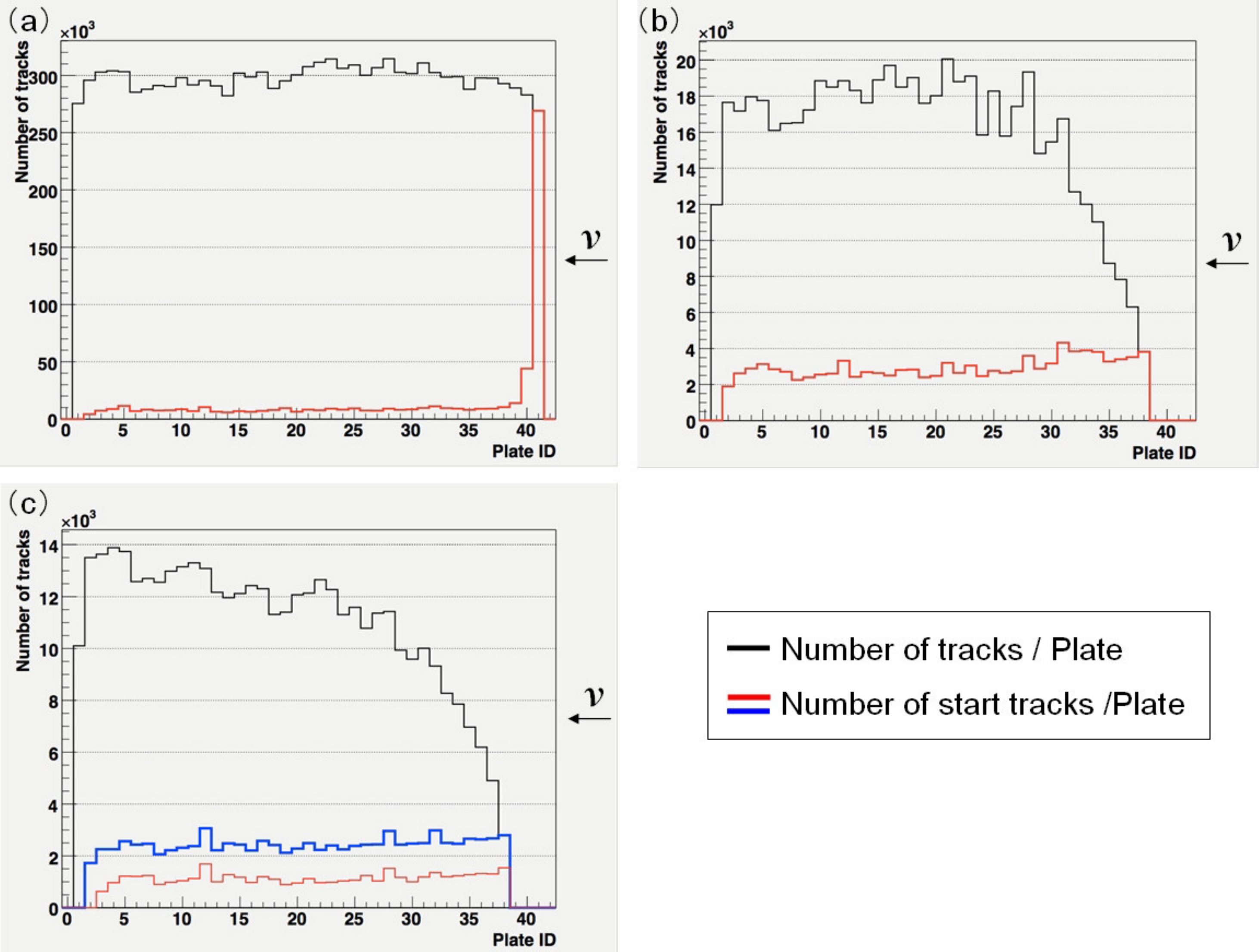}
\caption{(a): Number of tracks (black) and starting tracks (red) at each plate after fiducial selections. (b): Number of tracks (black) and starting tracks (red) at each plate after veto selections. (c): Number of tracks (black), number of starting tracks (blue), and number of starting tracks that have at least three segments (red) at each plate after final track selection.}
\label{ntrk}
\end{center}
\end{figure}

\begin{figure}[htbp]
\begin{center}
\includegraphics[clip, width=12.0cm]{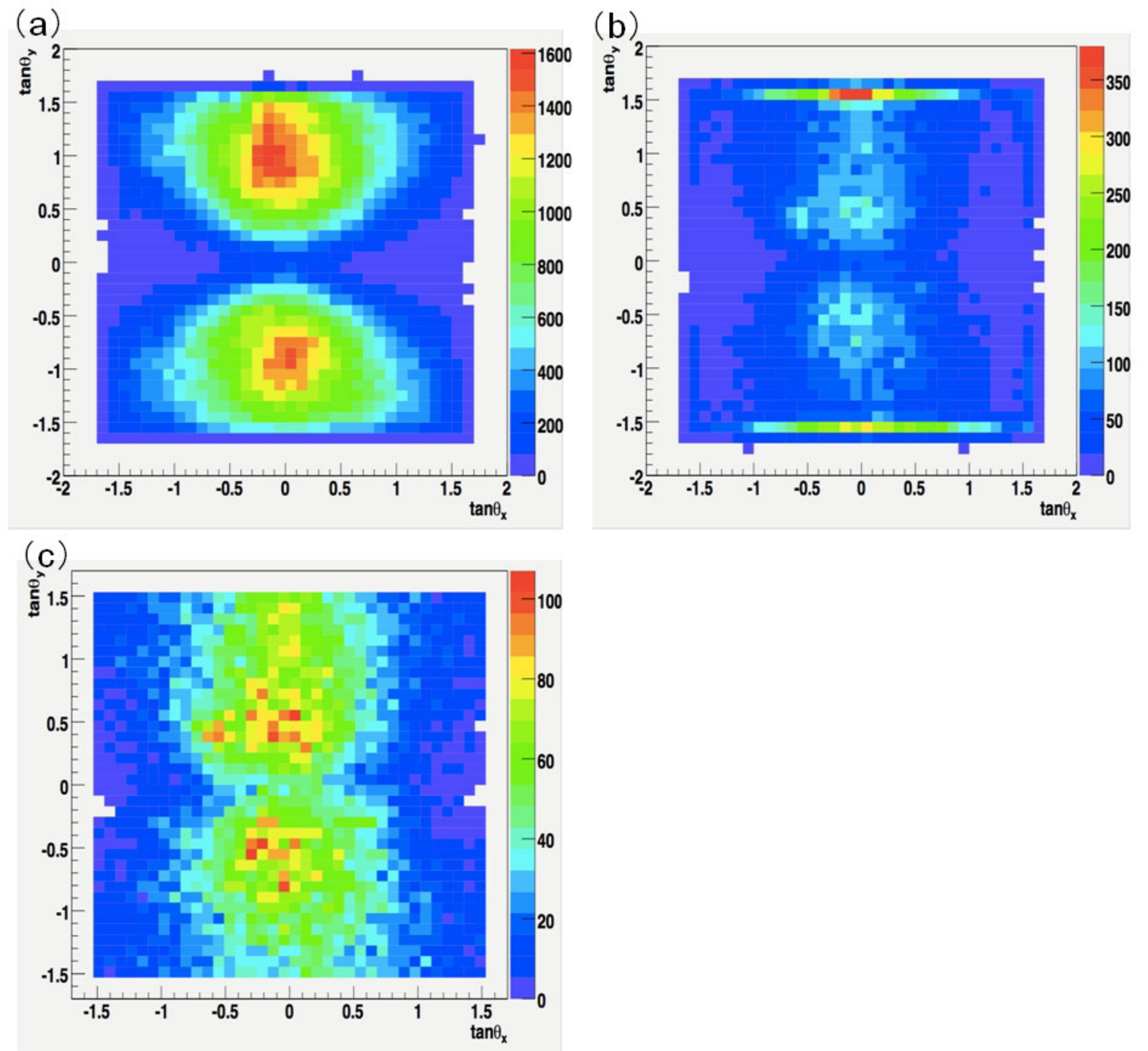}
\caption{(a): Angle distribution of penetrate and edge out tracks. (b): Angle distribution of remaining tracks after penetrate and edge out track cut. (c): Angle distribution of finally selected tracks for neutrino event search.}
\label{axay}
\end{center}
\end{figure}

\begin{enumerate}
\item Fiducial area selection : the acceptance of the average track angle is less than $|tan\theta|$ = 1.6 and the searched position area is set at $x$: 10--120 mm and $y$: 6--76 mm to avoid the bad film condition area (thin thickness of emulsion layer). The plate ID of the most upstream and downstream plates in the neutrino beam direction is given as PL41 and PL01, respectively. The number of tracks and the number of tracks that just start and connect to downstream plates after fiducial selections at each plate is described in Fig.\ref{ntrk}-(a). As shown by the red line in Fig.\ref{ntrk}-(a), there are many tracks at the upstream of the detector that are identified as penetrating tracks (cosmic ray muons or muons produced from outside of the detector).
\item Penetrate track veto : tracks that have at least one segment on three plates at the most upstream position of the ECC (PL41, PL40, PL39) are rejected as penetrate tracks.
\item Edge out track veto : tracks are rejected as edge out tracks in the case that the position extrapolated from the most upstream plate to its upstream plate using their position and angle is out of the fiducial area. The number of tracks and the number of starting tracks after veto selections at each plate is described in Fig.\ref{ntrk}-(b). The angle distribution of penetrate and edge out tracks is shown in Fig.\ref{axay}-(a). This is consistent with cosmic rays from the sky. Figure\ref{axay}-(b) shows the track angle distribution after (1)--(3) selections. The low-energy track or the electron/positron from gamma-ray or low efficiency tail from large angle tracks still remain as background for starting tracks in the ECC.
\item Final track selection for vertex search : the tracks that have at least three segments and for which the track angle is less than $|tan\theta|$ = 1.5 ($|\theta| \sim$ 56 degree) at the most upstream plate are selected to avoid accidental disconnecting tracks in the large-angle region and low reliability tracks. Finally, 41087 tracks are selected as the candidate tracks from neutrino interactions. The number of tracks and the number of starting tracks after these selections at each plate is described in Fig.\ref{ntrk}-(c). Additionally, the track angle distribution after this selection is shown in Fig.\ref{axay}-(c).
\end{enumerate}

\clearpage

\subsubsection{Event Selection}
\begin{figure}[h]
\begin{center}
\includegraphics[clip, width=17.0cm]{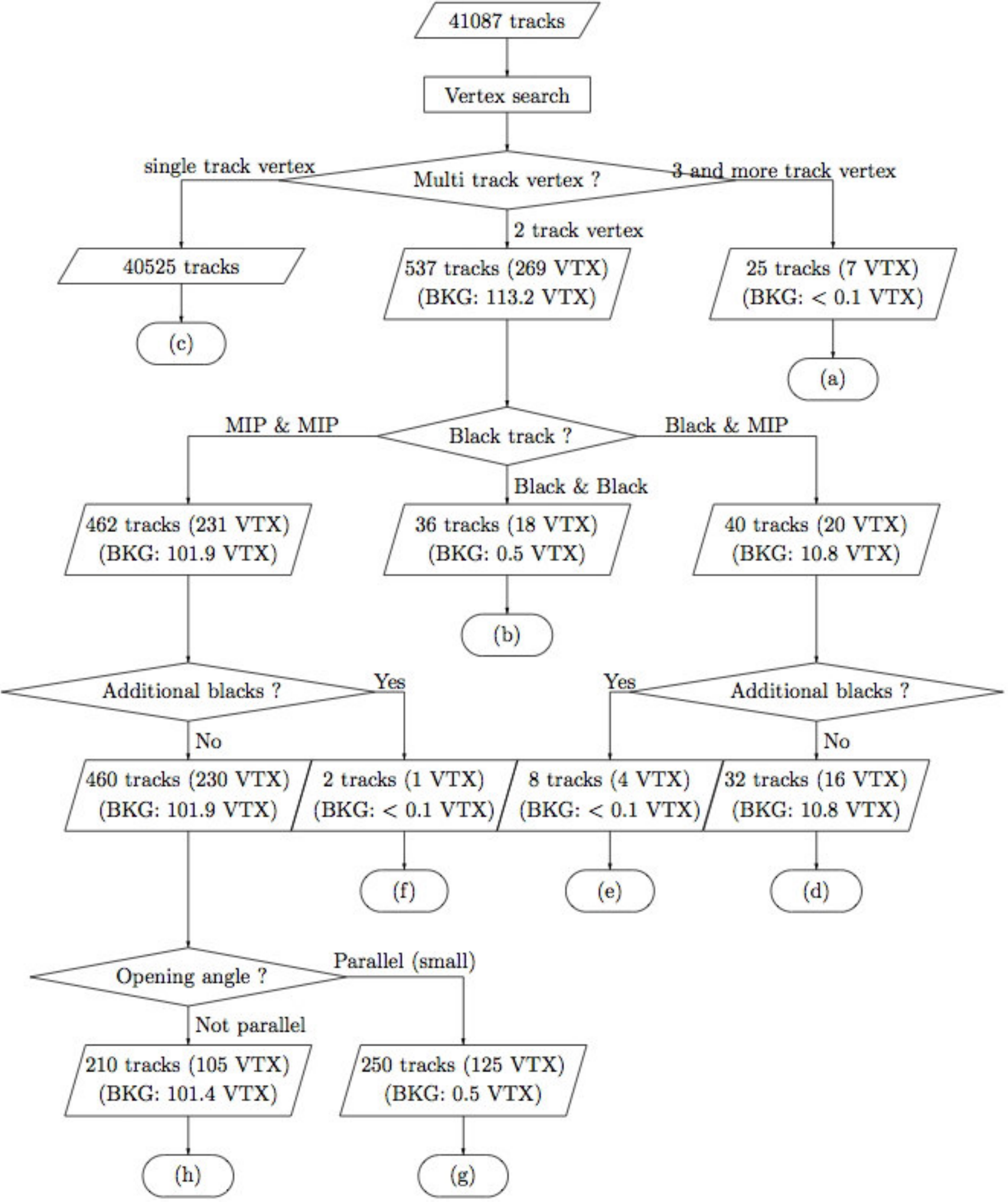}
\caption{Flow chart for event selection}
\label{flow}
\end{center}
\end{figure}

The event selection is performed with the following procedure. The events are then categorized into the types of (a)--(h), as shown in Fig.\ref{flow}. 

\begin{figure}[t]
\begin{center}
\includegraphics[clip, width=13.0cm]{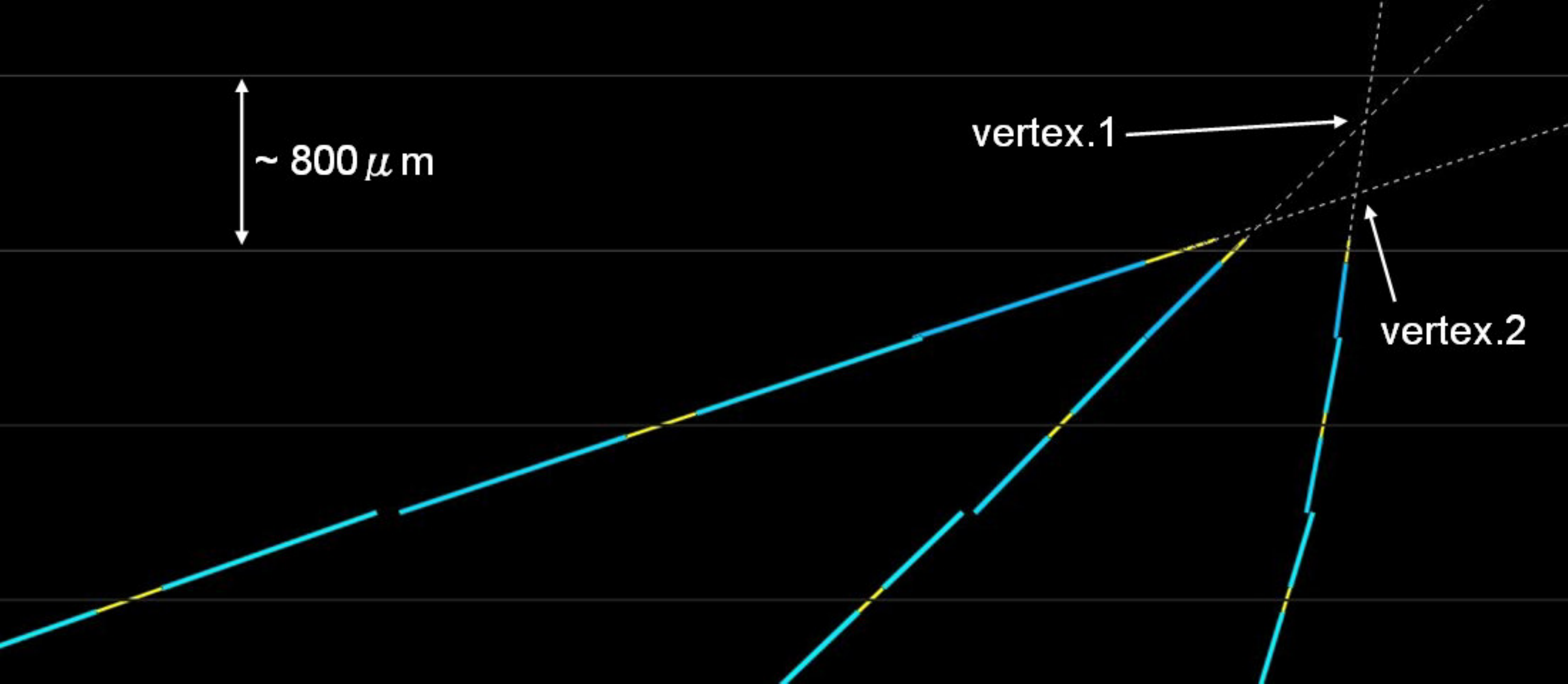}
\caption{ Two vertices are reconstructed by three tracks}
\label{3trk2vtx}
\end{center}
\end{figure}

\begin{figure}[b]
\begin{center}
\includegraphics[clip, width=15.0cm]{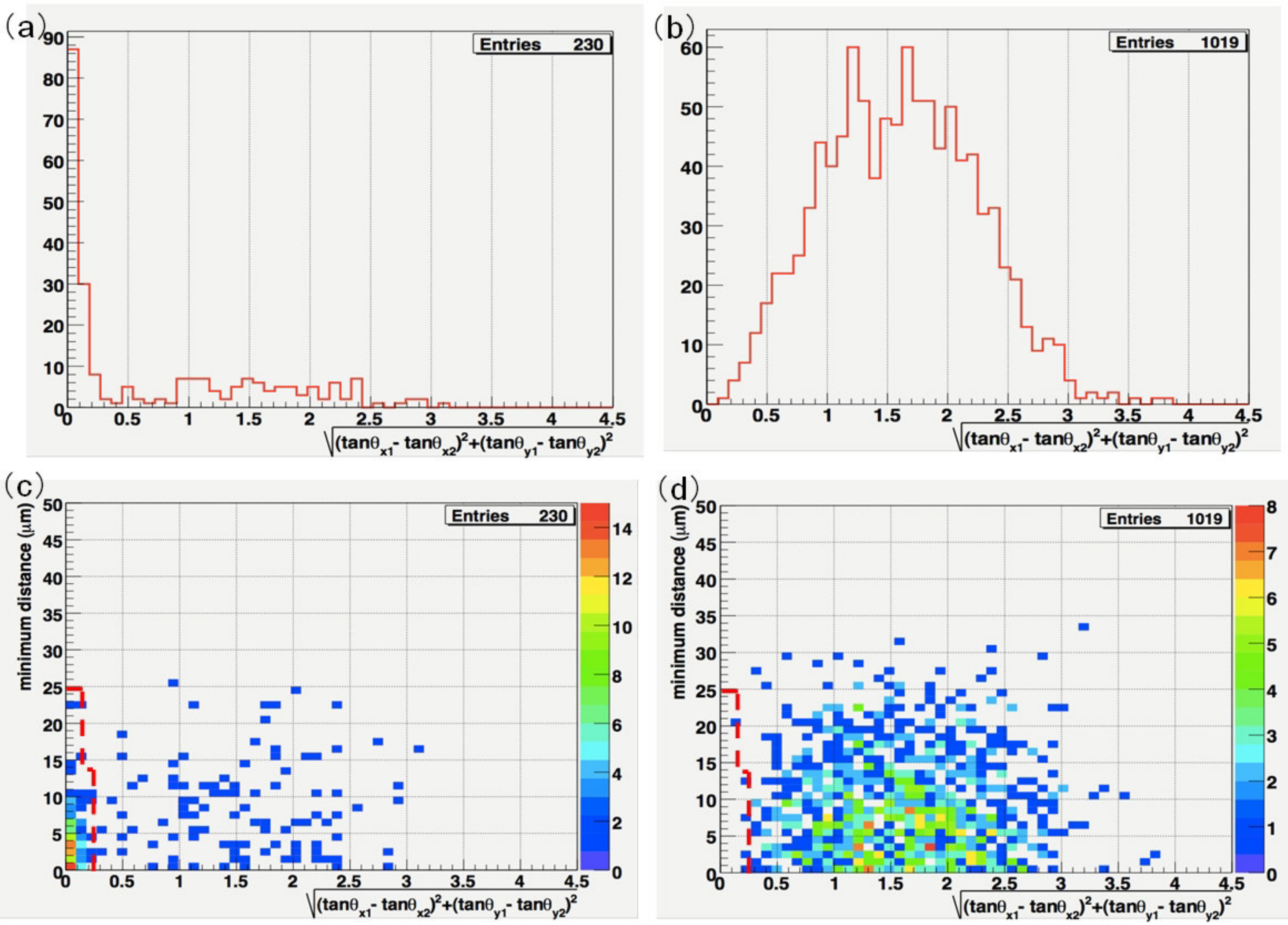}
\caption{Opening angle between two tracks in real data (a) and 10 times of the estimated background data (b). The opening angle vs. minimum distance between two tracks in real data (c) and 10 times of the estimated background data (d). Gamma-ray events are selected within the dotted red line in (c) and (d). }
\label{gammasel}
\end{center}
\end{figure}

\begin{enumerate}
\item Vertex Search: the most upstream base tracks converge at a point within 3 ${\sigma}$ position difference that is calculated from their angular accuracy and the MCS assumed the momentum of 400 MeV/c. For example, the position allowance for the minimum distance between two tracks (tan$\theta$ = 0.5, respectively) is approximately 20 $\mu$m at the level of 500 $\mu$m of iron above the surface of emulsion film. In this search, the tracks that emit to the forward direction are selected as vertex tracks.

Most of starting tracks are the accidental disconnecting tracks or the stopped cosmic-ray tracks from the back of the ECC. So the positions of these tracks are random and fake vertices are created with their accidental chance coincidence. Such background vertices can be emulated by shuffling only track positions randomly in each plate and making vertices. Repeating 10 times procedure, the number of background is estimated as the average value in Fig.\ref{flow}.

\item Track Multiplicity: as a first step, reconstructed vertices are categorized based on their track multiplicity. The three categories are "three and more tracks", "two tracks" and "single track". This "three and more tracks" is categorized as Category-(a). 

In Category-(a), seven vertices composed of 25 tracks are found and the estimated number of background vertex is less than 0.1. Thus, the vertices of Category-(a) are real vertex candidates because of low background. In the two-track vertex category, there are 269 vertices composed of 537 tracks including one duplicated case, as shown in Fig.\ref{3trk2vtx}, corresponding to 112.6 background vertices. Then, 40525 tracks are categorized as single track vertex event (Category-(c)). Single track vertices are mainly considered to be disconnecting of low-energy cosmic ray muons or one of an electron/positron pair production, as shown in Fig.\ref{axay}-(c). 
\item Track Blackness: two-track vertices are categorized according to track blackness. Black tracks are defined as tracks whose average VPH is over 120, and minimum ionizing particle (MIP) tracks are defined as tracks whose average VPH is less than 120, as shown in Fig.\ref{pid}. When two-track vertices are reconstructed by only black tracks, they belong to Category-(b). In Category-(b), 18 vertices composed of 36 tracks are found, corresponding to 0.5 background vertices. Thus, they are also real vertex candidates. When one black track and one MIP track made a vertex, there are 20 vertices with 40 tracks and the number of the estimated background is 10.8. 
\item Additional Black Search: the short black tracks made by only one or two plates are added as vertex tracks. This procedure is done by the short black tracks both the forward and backward directions. When the short black track is searched around 20 vertices that are reconstructed by one black track and one MIP track, four new vertices with multiple black tracks are found with less than 0.1 background (Category-(e)). Other vertices are stored in Category-(d). 231 two track vertices that are reconstructed by only MIPs are also attached to additional black tracks. One new vertex with two MIPs and one short black track is found. These vertices are stored in Category-(f).
\item Track Parallelity: Fig.\ref{gammasel}-(a) shows the opening angle of 230 MIP vertices with no additional black tracks. The opening angle of ten times of background data is shown in Fig.\ref{gammasel}-(b). According to the comparison between real and background data, an excess of small opening angle is found in real data. They are almost parallel tracks, so are explained as the electron/positron pair production from gamma-ray. Gamma-ray candidate events are selected by using their opening angle and minimum distance of vertex tracks as shown in Fig.\ref{gammasel}-(c). 125 vertices are selected as gamma-ray events and the estimated number of background is 0.5 as shown in Fig.\ref{gammasel}-(d). Gamma-ray events are stored in Category-(g) and the other vertices are stored in Category-(h). 
\end{enumerate}

Through the event selections, Categories (a), (b), (e), (f), and (g) are selected as a vertex events with high reliability. The 125 events in Category-(g) are gamma-ray candidates because they have no black track and their opening angle is small. The 18 vertices in Category-(b) are reconstructed by only black tracks, so it is considered that these events are from neutron interactions or a part of neutrino neutral current interactions. The neutrino candidate events are stored in Categories (a), (e), and (f). The 12 vertices in these categories have some MIP and black tracks.

On the other hand, there may be some neutrino events in Category-(d) because the number of vertex signals is 5.2 ($= 16 - 10.8$). The near vertices in Category-(h) are accidental chance coincidence of misconnected cosmic-ray muons or one of electron/positron pair. It is considered that they are same group as Category-(c).

\vspace*{\stretch{1}}

\vspace{\stretch{1}}

\clearpage

\clearpage

\subsection{Gamma-ray events}

\begin{figure}[ht]
\begin{center}
\includegraphics[clip, width=12.5cm]{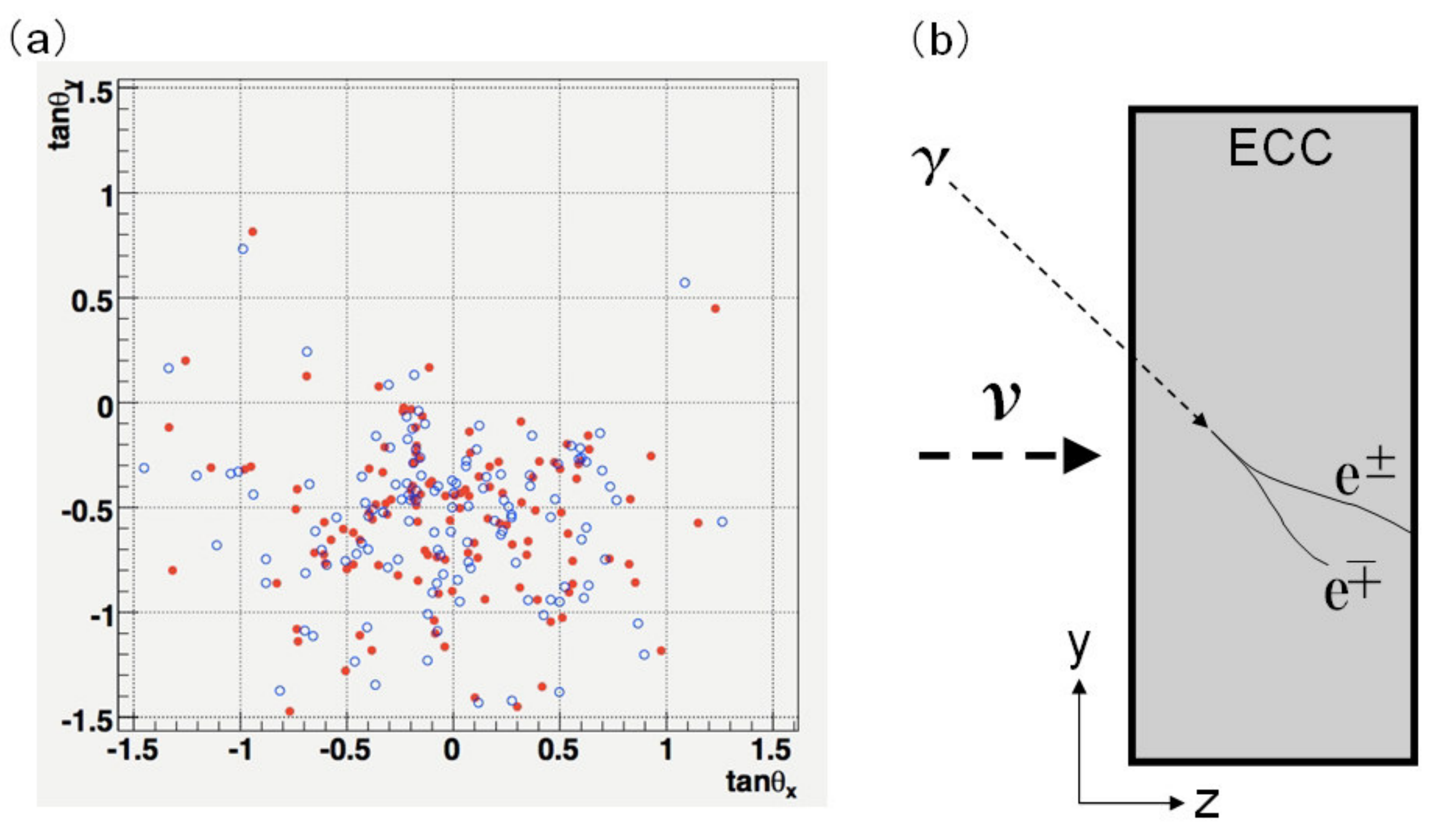}
\caption{Angle distribution of electron/positron (red or blue) from gamma-ray.}
\label{gammaang}
\end{center}
\end{figure}

\begin{figure}[b]
\begin{center}
\includegraphics[clip, width=8.5cm]{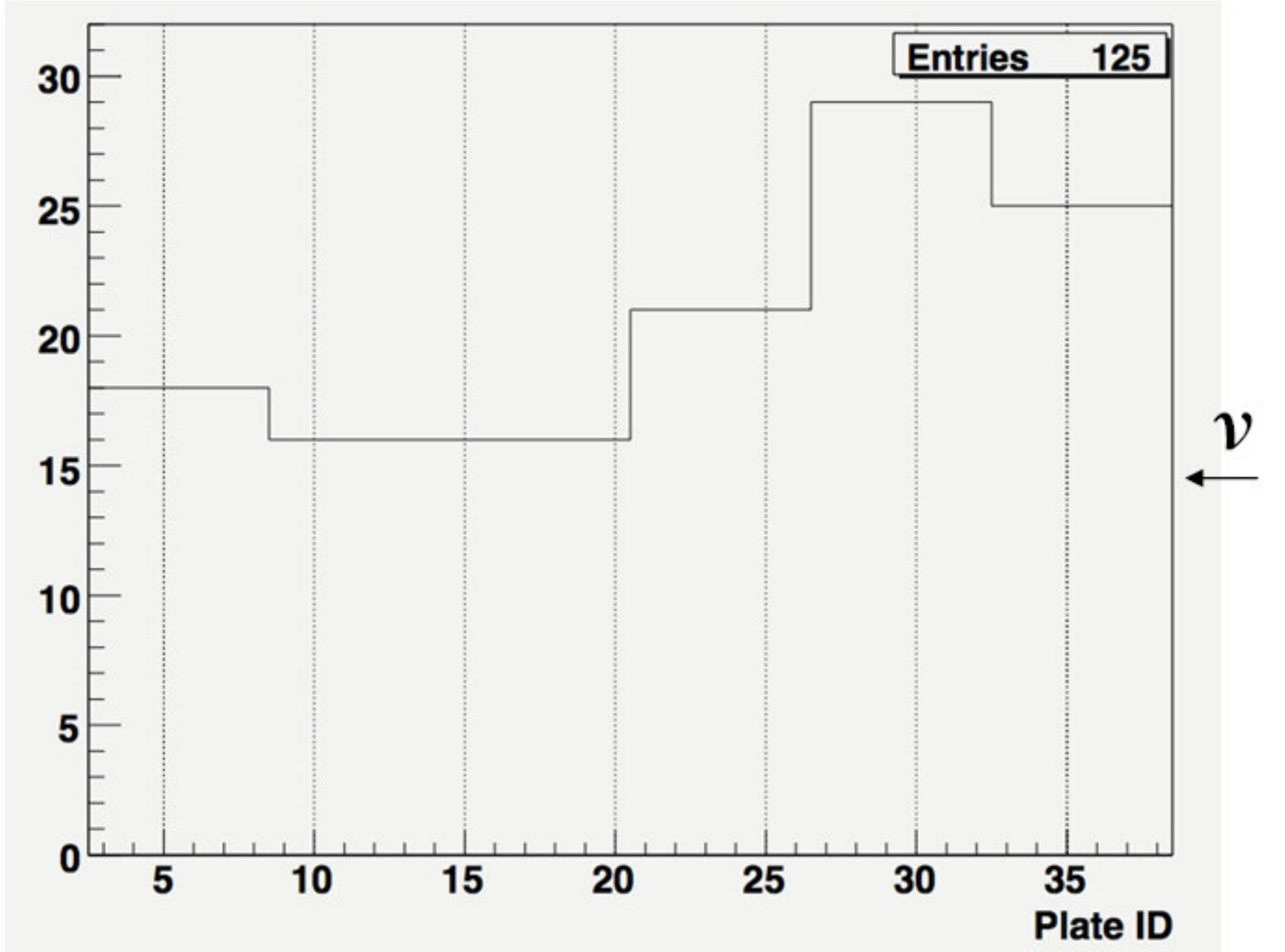}
\caption{Converted plate distribution of gamma-rays.}
\label{gammapl}
\end{center}
\end{figure}

\begin{figure}[htbp]
\begin{center}
\includegraphics[clip, width=15.0cm]{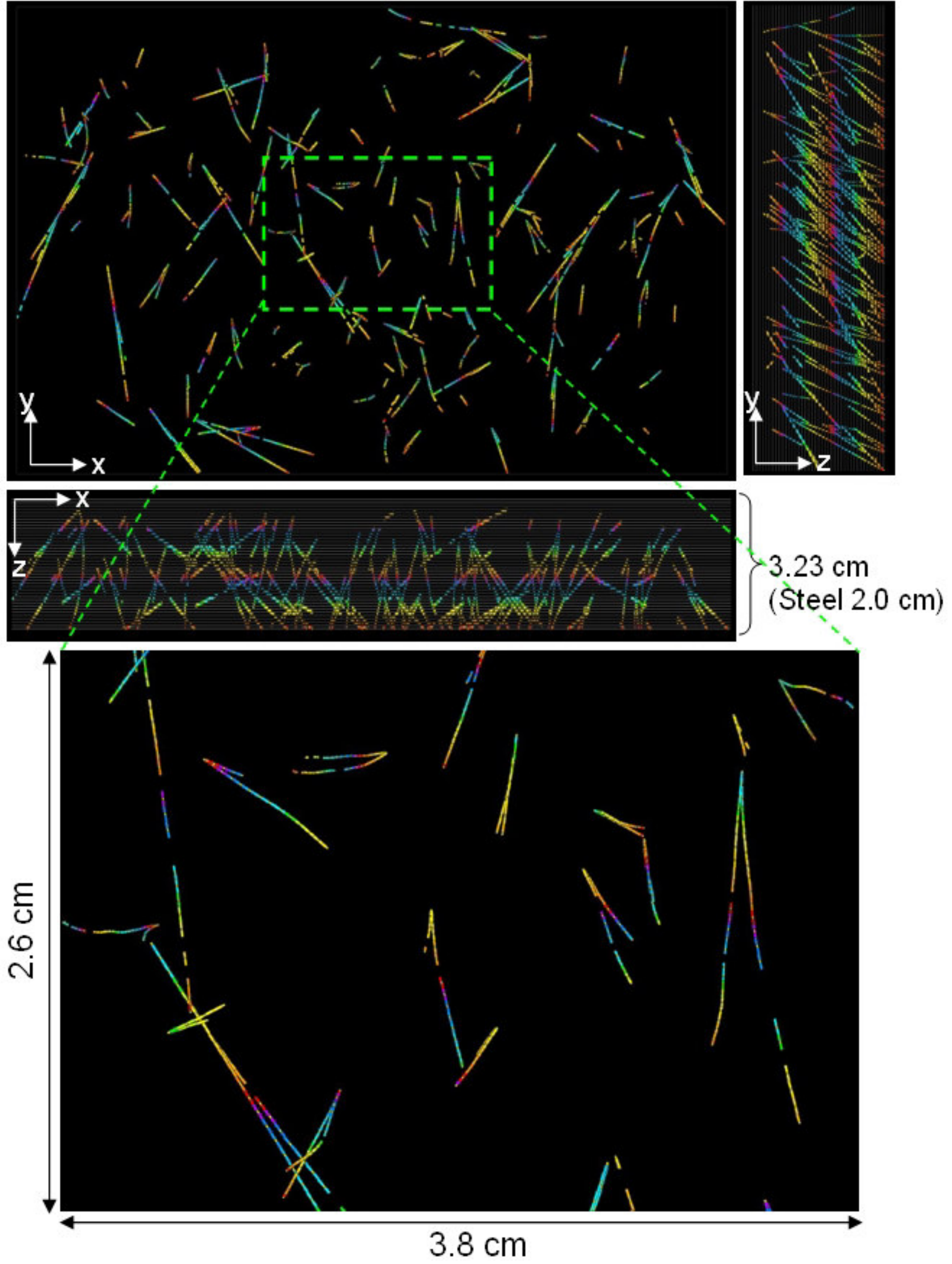}
\caption{The 3 dimensional (x-y, x-z, y-z) and zooming view. Neutrino beam direction is same as z axis.}
\label{gammamap}
\end{center}
\end{figure}

In this section, Category-(g), or gamma-ray candidate events, will be discussed. Only forward going gamma-ray is reconstructed in this category. The angle distribution of gamma-ray candidate events is shown in Fig.\ref{gammaang}-(a). The incident direction of tracks from gamma-ray is forward and upper side and it is consistent with the direction of cosmic-rays (Fig.\ref{gammaang}-(b)). Additionally, the number of converted gamma-ray is high in the upstream region of the detector, as shown in Fig.\ref{gammapl}. This is evidence that the gamma-rays come from outside the detector. The detected electron-positron pairs from gamma-ray in the ECC is shown in Fig.\ref{gammamap}. It is considered that cosmic gamma-rays from the backward and upper side directions make the single-track vertex in the Category-(c) and background vertices.

\clearpage

\clearpage

\subsection{Black track vertex events}

\begin{figure}[htbp]
\begin{center}
\includegraphics[clip, width=15.0cm]{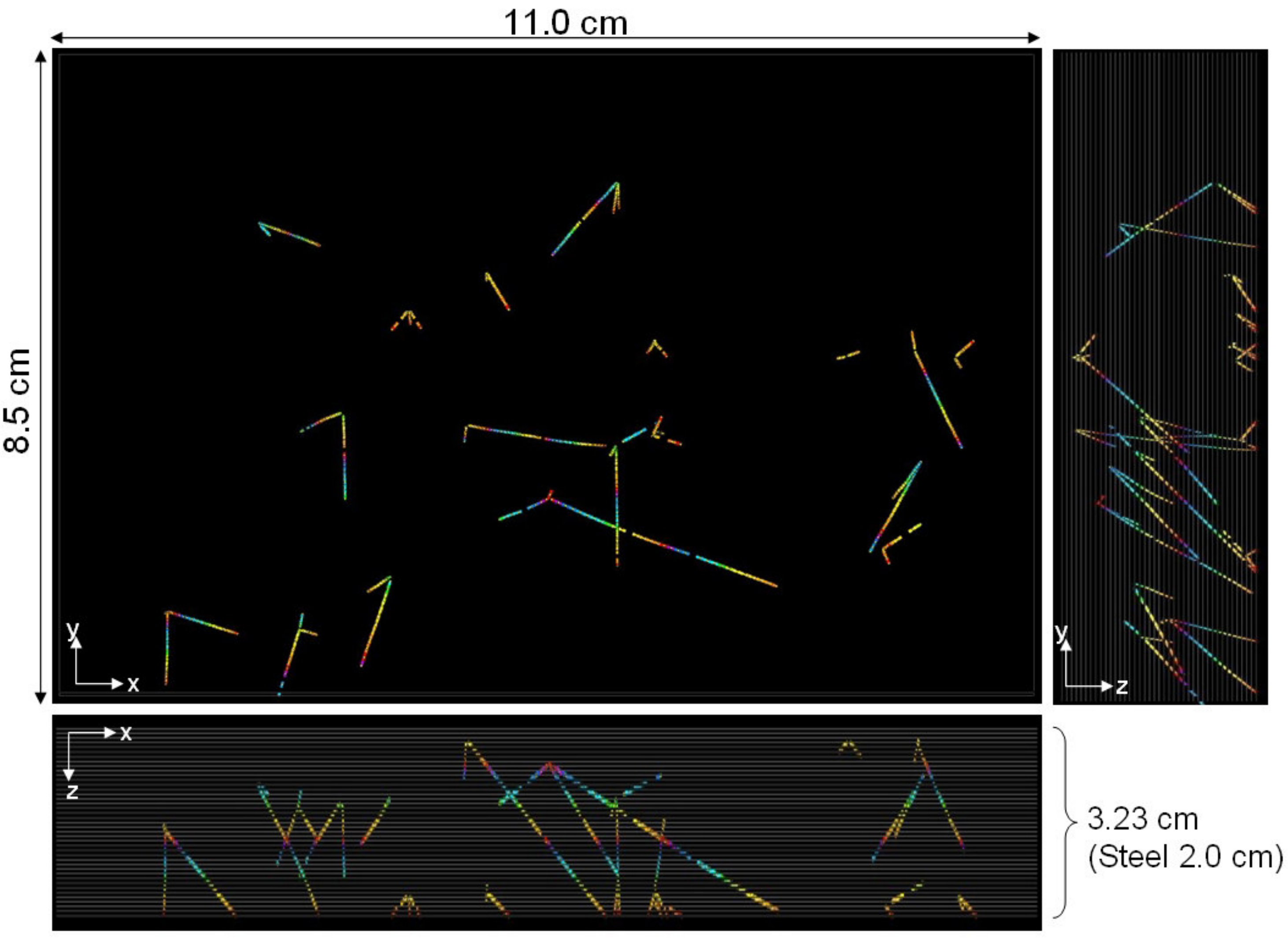}
\caption{Three-dimensional view of black vertex events. Neutrino beam direction is parallel to the $z$-axis.}
\label{blackmap}
\end{center}
\end{figure}

\begin{figure}[htbp]
\begin{center}
\includegraphics[clip, width=15.0cm]{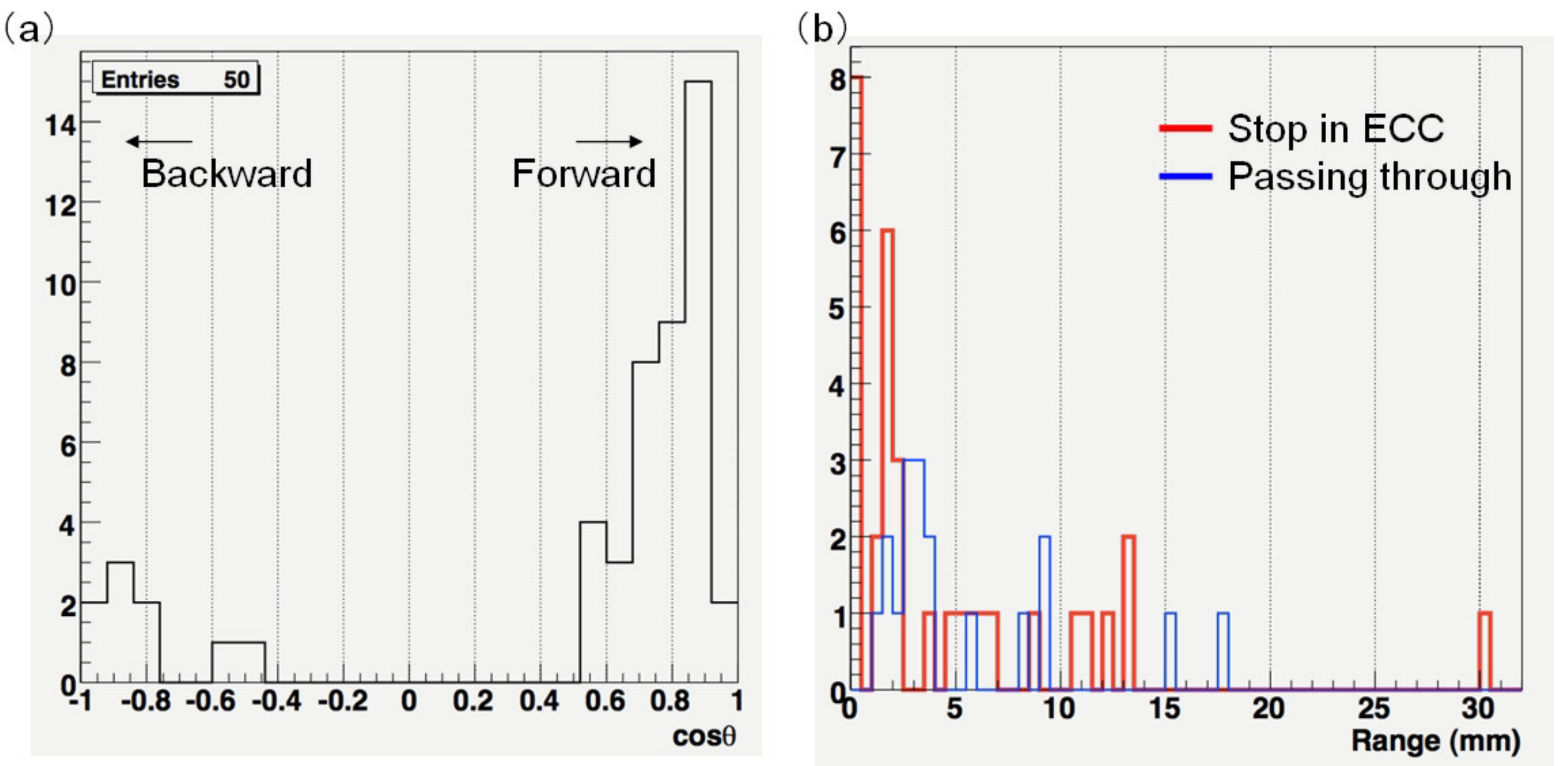}
\caption{Emission angle distribution (a) and the range distribution (b) of tracks in black vertex event.}
\label{black}
\end{center}
\end{figure}

In this section, a discussion of 18 found vertices that are reconstructed by only black tracks in Category-(b) will be presented. The chance coincidence background is estimated as 0.5 events. Therefore, almost events are real vertices. These events are considered as the neutron interactions or a part of neutrino neutral current interactions with no MIP tracks. Additional short or backward black tracks attached to these vertices were also searched for. As shown in table.\ref{blackvtx}, new black tracks were found in more than half of events. The event feature is shown in Fig.\ref{blackmap}. The emission angle and range distributions are shown in Fig.\ref{black}-(a) and (b), respectively. Under the  current selection criteria, the existence of two black tracks that are emitted to the forward direction is required. Thus, this becomes one of the biases to drop neutron events from the backward direction.

\begin{table}[!h]
\caption{Black track vertices}
\label{blackvtx}
\centering
\begin{tabular}{|c||c||c|}
\hline
2 track vertex & 3 track vertex & 4 track vertex\\ 
\hline
7 & 8 & 3\\
\hline
\end{tabular}
\end{table}

\clearpage

\subsection{Neutrino Candidate Events}
The other converged tracks, for example, low-energy tracks, tracks emitted to the backward direction, or electron/positron pair from vertices are searched for in the 12 vertices found in Categories (a), (e), and (f). The detected particle momenta are then measured with MCS and the timing information is obtained with the emulsion shifter to connect the INGRID and identify muons \cite{t60shifter}. In this section, the event summary is shown in table \ref{summary} and then some of the detailed event features will be described. 

\begin{table}[!h]
\caption{Event summary}
\label{summary}
\centering
\footnotesize
\begin{tabular}{|c|c|c|c|c|c|}
\hline
\shortstack{Event \\ number} & VTX Plate & material & \shortstack{Multiplicity \\ (MIP)} & \shortstack{Multiplicity \\ (Black)} & \shortstack{Multiplicity \\ (e$^+$e$^-$)} \\ 
\hline
1 & 8 & iron & 5 & 0 & 3 \\
2 & 13 & iron & 2 & 1 & 0 \\
3 & 14 & iron & 3 & 2 & 0 \\
4 & 23 & iron & 2 & 2 & 0 \\
5 & 26 & iron & 1 & 7 & 0 \\
6 & 26 & iron & 1 & 3 & 1 \\
7 & 27 & emulsion & 2 & 2 & 1 \\
8 & 28 & emulsion & 3 & 7 & 0 \\
9 & 30 & iron & 1 & 2 & 0 \\
10 & 31 & iron & 1 & 2 & 1 \\
11 & 33 & iron & 2 & 2 & 1 \\
12 & 36 & iron & 1 & 2 & 0 \\

\hline
\end{tabular}
\end{table}

\clearpage

\begin{figure}[ht]
\begin{center}
\includegraphics[clip, width=15.0cm]{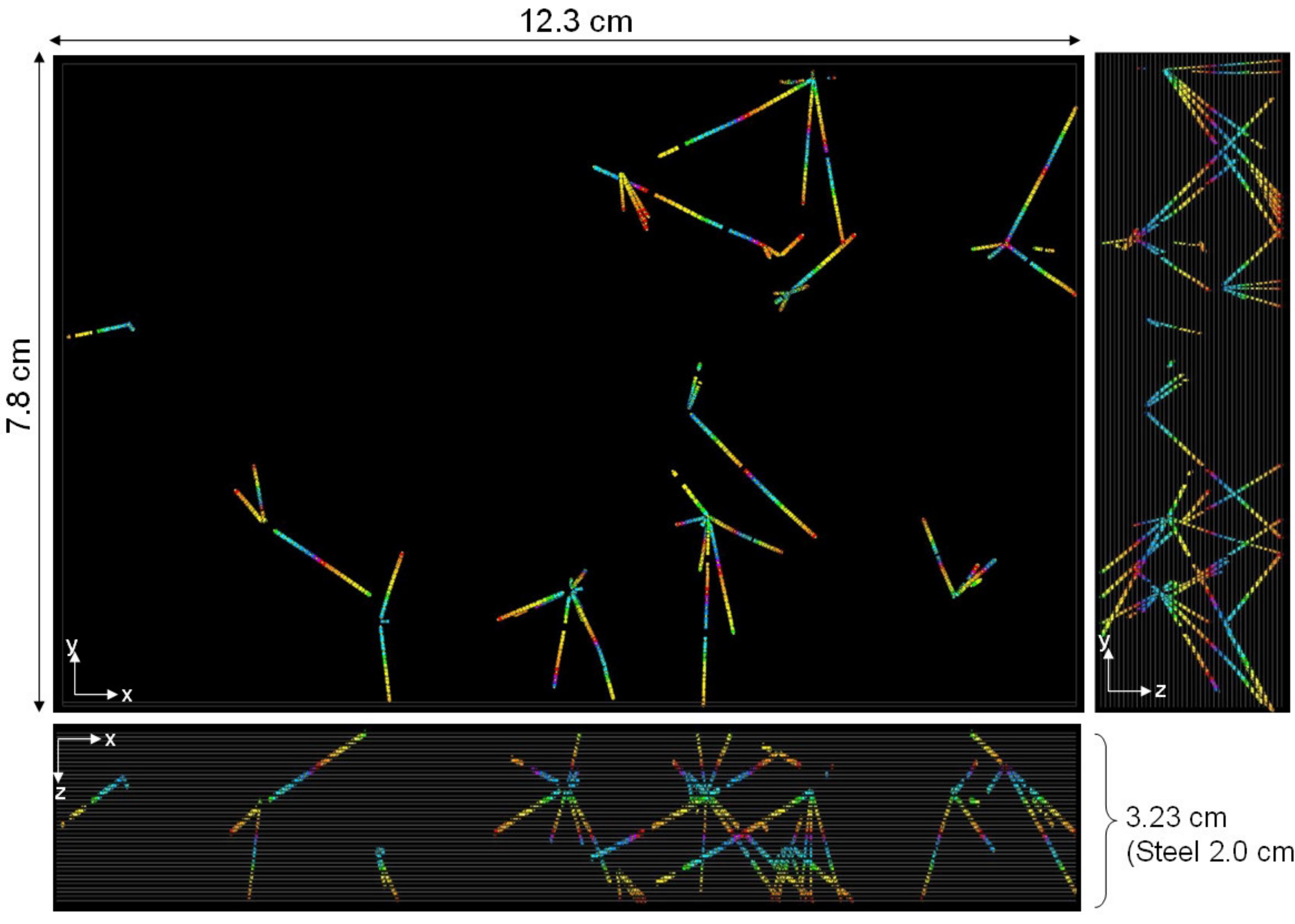}
\caption{Three-dimensional view of neutrino candidate events. Neutrino beam direction is parallel to the $z$-axis.}
\label{nuevent}
\end{center}
\end{figure}

\begin{figure}[ht]
\begin{center}
\includegraphics[clip, width=14.0cm]{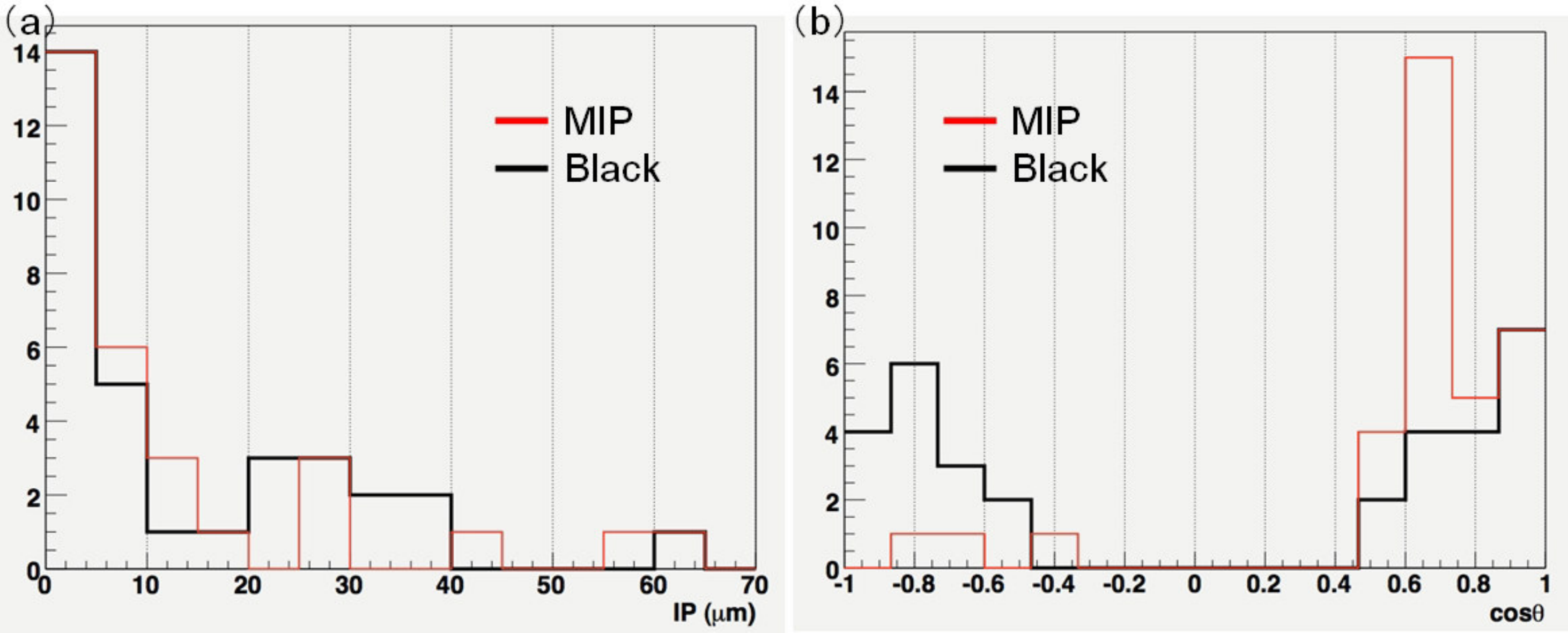}
\caption{Impact Parameter and Emission angle for MIP and black}
\label{ipcos}
\end{center}
\end{figure}

The position distribution of 12 vertices is shown in the Fig.\ref{nuevent}. The impact parameter (IP) and emission angle for each MIP and black tracks is also shown in the Fig.\ref{ipcos}.

\clearpage

\subsubsection{Event.7}
\begin{figure}[ht]
\begin{center}
\includegraphics[clip, width=15.0cm]{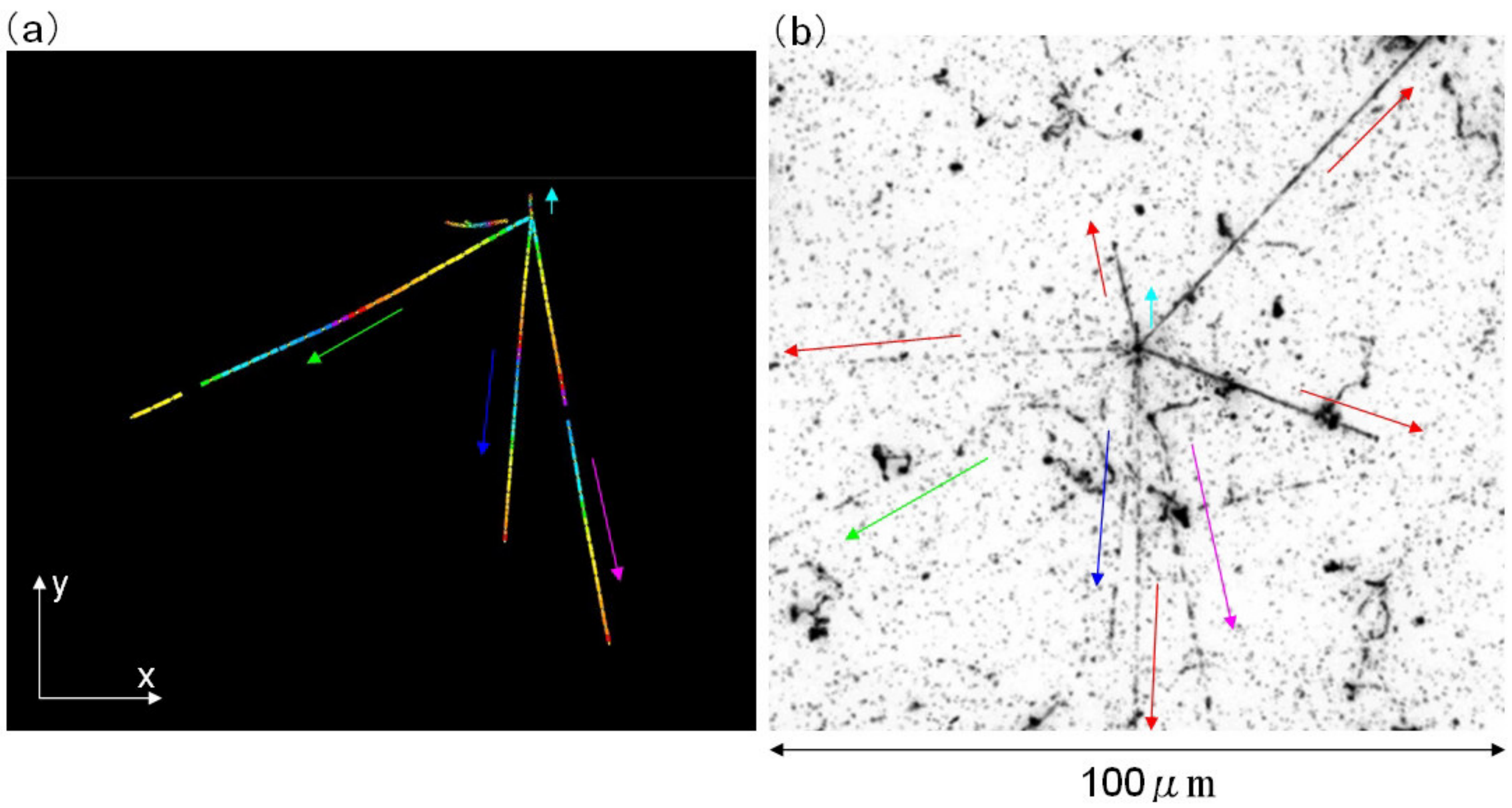}
\caption{Event.7}
\label{ev009z}
\end{center}
\end{figure}

\begin{figure}[ht]
\begin{center}
\includegraphics[clip, width=13.0cm]{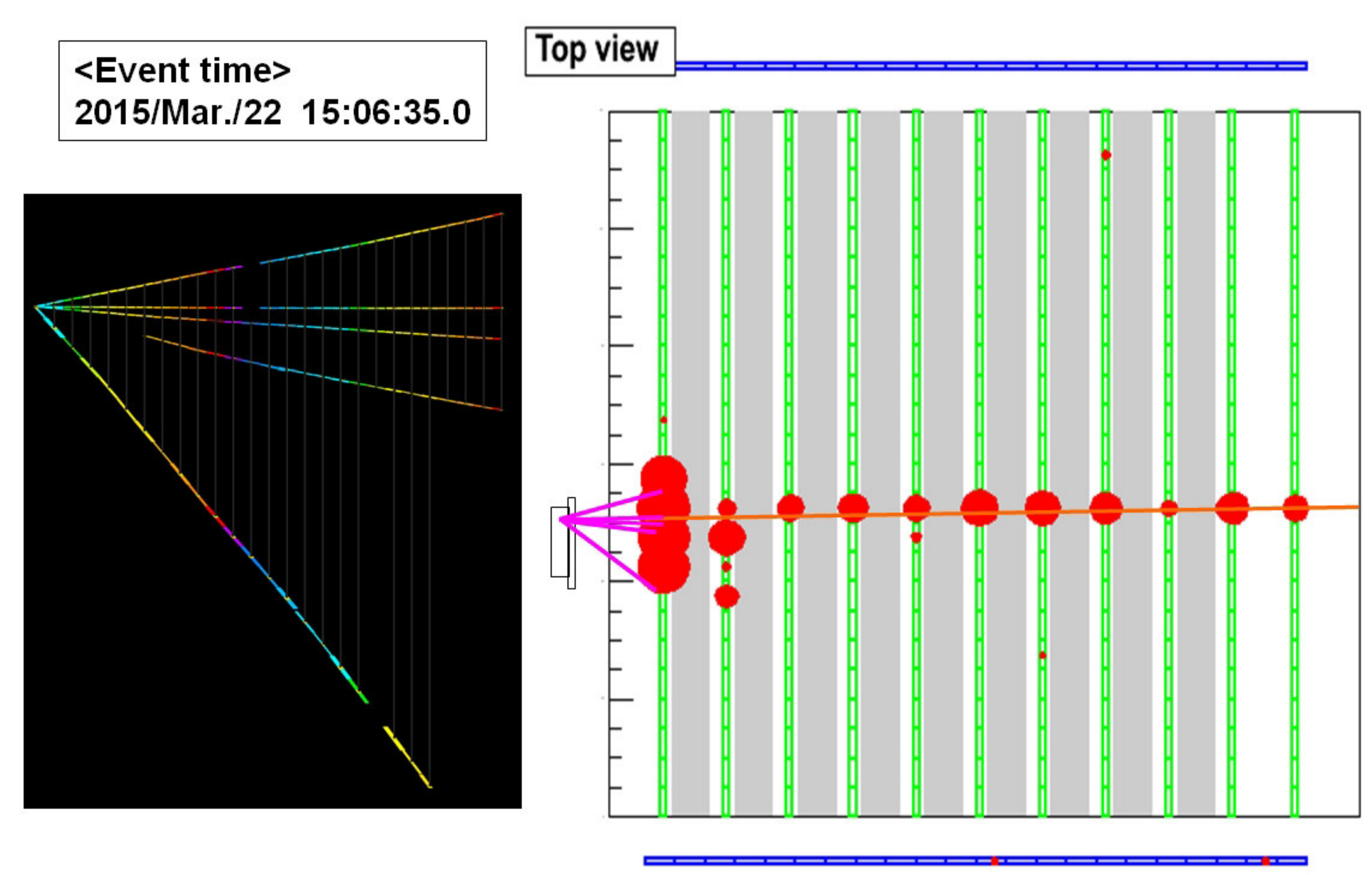}
\caption{Event.7: Hybrid analysis with INGRID}
\label{ev4top}
\end{center}
\end{figure}

\begin{figure}[ht]
\begin{center}
\includegraphics[clip, width=13.0cm]{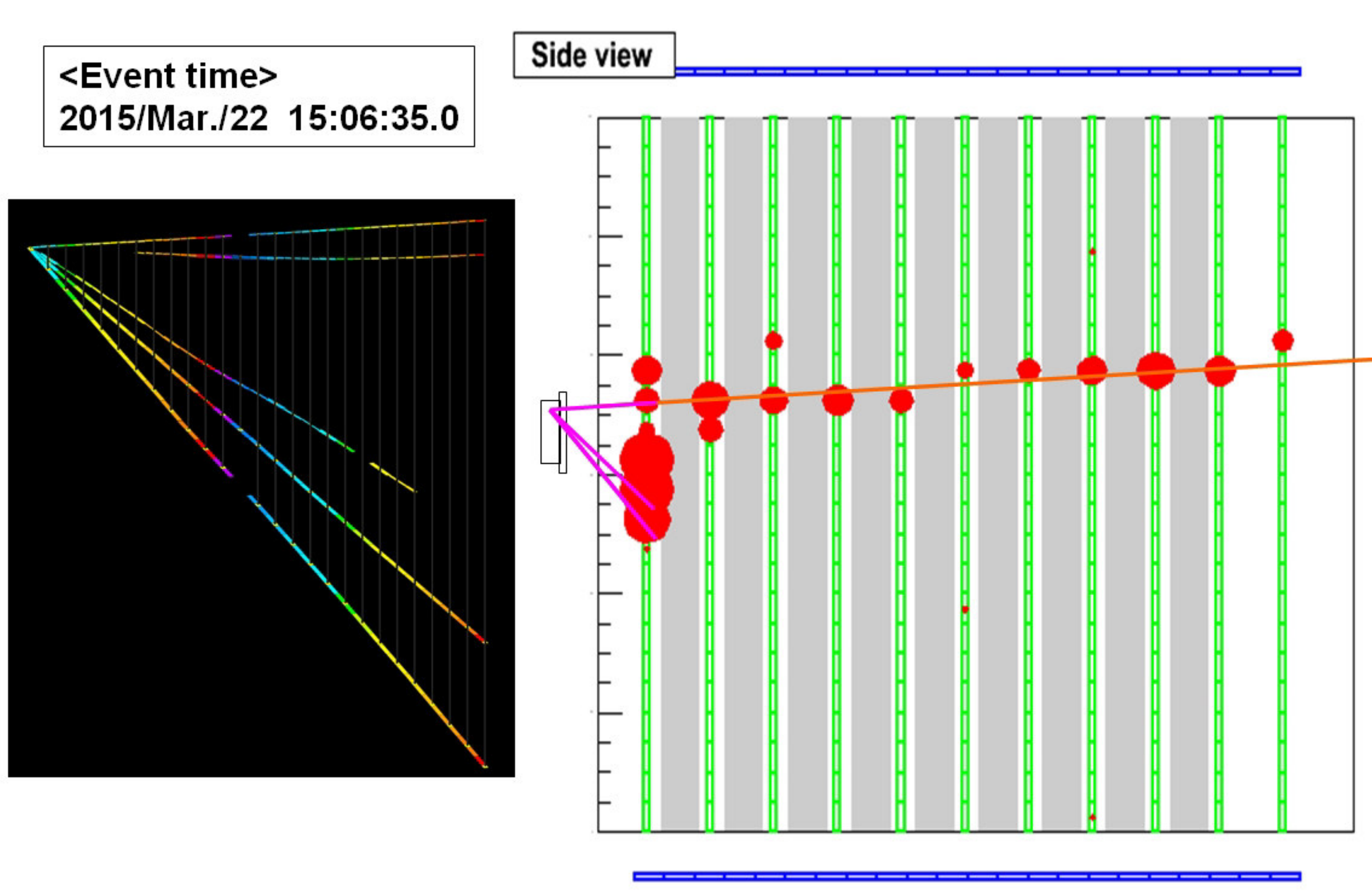}
\caption{Event.7: Hybrid analysis with INGRID}
\label{ev4side}
\end{center}
\end{figure}

The vertex is in emulsion region at plate.27 and its vertex point is $(x,y) = (96034.4$ $\mu$m, $74184.9$ $\mu$m$)$ in the ECC coordinate system (the origin (0.0 $\mu$m, 0.0 $\mu$m) is the corner at PL01). One electron/positron pair from a gamma-ray in Category-(g) points to this vertex point. The detailed track information is listed in the table \ref{vtxinf1}. There are two MIP tracks and two black tracks in the vertex (Fig.\ref{ev009z}-(a)). Also, this event was created at the emulsion region, so we could see more detail with microscope. As shown in Fig.\ref{ev009z}-(b), the additional very large angle tracks and black tracks, shown as red arrows, were observed. Additionally, this event was given the time stamp by analyzing the emulsion shifter, then the timing matching between the emulsion detector and INGRID succeeded uniquely. As shown in Fig.\ref{ev4top} and Fig.\ref{ev4side}, a muon track is found and the event feature is well matched. Furthermore, the measured track momenta in ECC are consistent with the measured track ranges in INGRID. The momenta of muon candidate track is also enough to pass through INGRID.

\begin{table}[!h]
\caption{Track information in Event.7}
\label{vtxinf1}
\centering
\begin{tabular}{|c|c|c|c|c|c|c|c|c|}
\hline
PL & segment id & ${tan\theta_x}$ & ${tan\theta_y}$ & \shortstack{IP \\ (${\mu}$ m)} & PH$_{AV.}$ & VPH$_{AV.}$ & \shortstack{p${\beta}$ \\ (GeV/c)} & PID \\ 
\hline

27 & 608427 &  0.2064 & -1.0459 &  0.7 & 29.5 & 182.8 & 0.43$^{+0.07}_{-0.07}$ & proton \\
27 & 608429 & -0.0661 & -0.7599 &  0.6 & 27.7 & 122.3 & 0.46$^{+0.07}_{-0.07}$ & proton \\
27 & 608428 & -0.0059 &  0.0540 &  0.7 & 30.2 & 118.3 & 4.00$^{+\infty}_{-1.75}$ & muon \\
27 & 608433 & -0.9756 & -0.5322 &  0.9 & 24.7 & 60.8 & 0.21$^{+0.03}_{-0.03}$ & MIP \\
21 & 586893 & -0.2365 & -0.0407 & 29.3 & 27.9 & 70.0 & 0.21$^{+0.02}_{-0.03}$ & e$^\pm$ \\
21 & 586883 & -0.2190 & -0.0666 & 159.3 & 27.7 & 72.1 & 0.08$^{+0.02}_{-0.02}$ & e$^\mp$ \\

\hline
\end{tabular}
\end{table}

\clearpage

\subsubsection{Event.3}
\begin{figure}[ht]
\begin{center}
\includegraphics[clip, width=13.0cm]{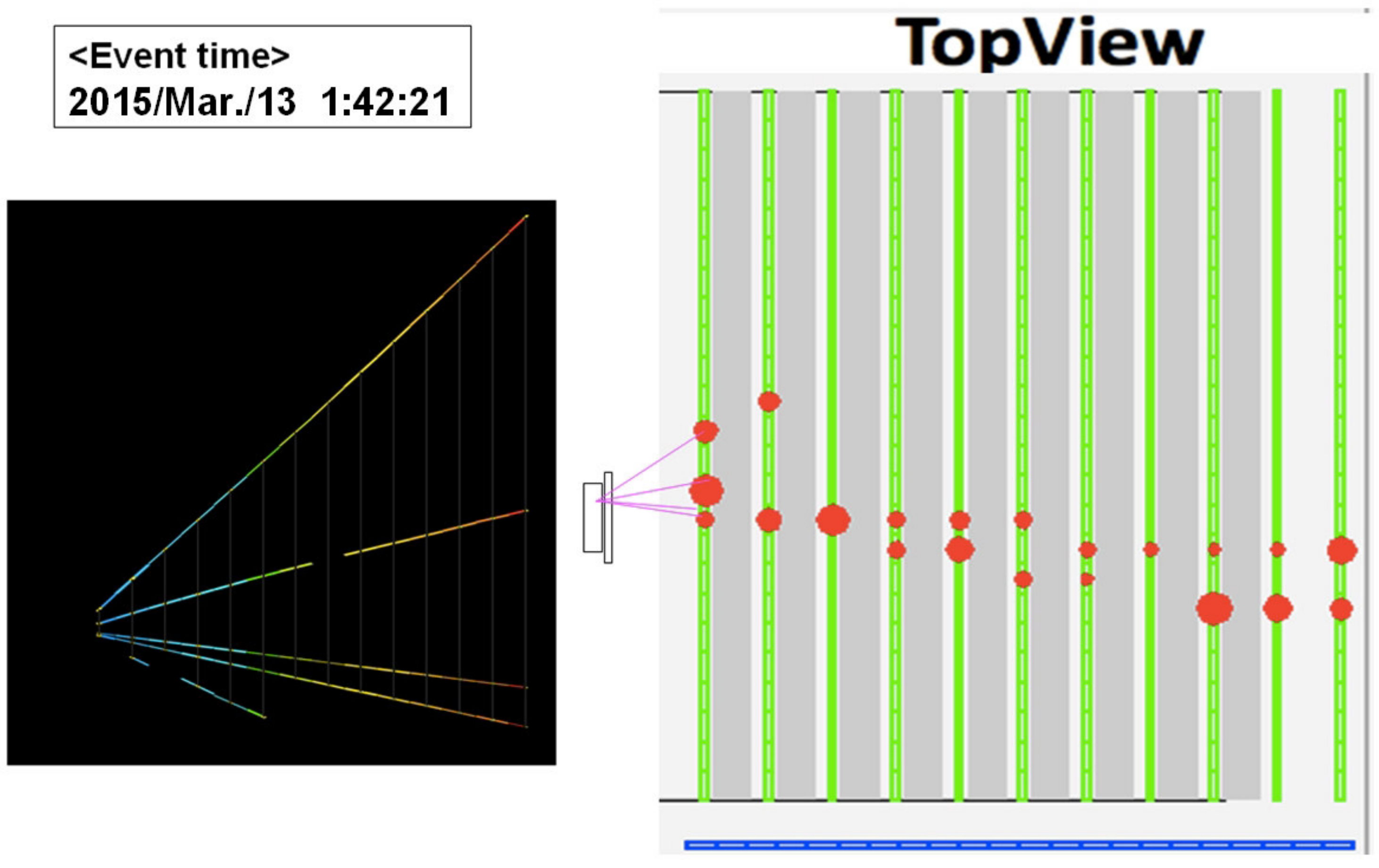}
\caption{Event.3: Hybrid analysis with INGRID}
\label{ev2top}
\end{center}
\end{figure}

\begin{figure}[ht]
\begin{center}
\includegraphics[clip, width=13.0cm]{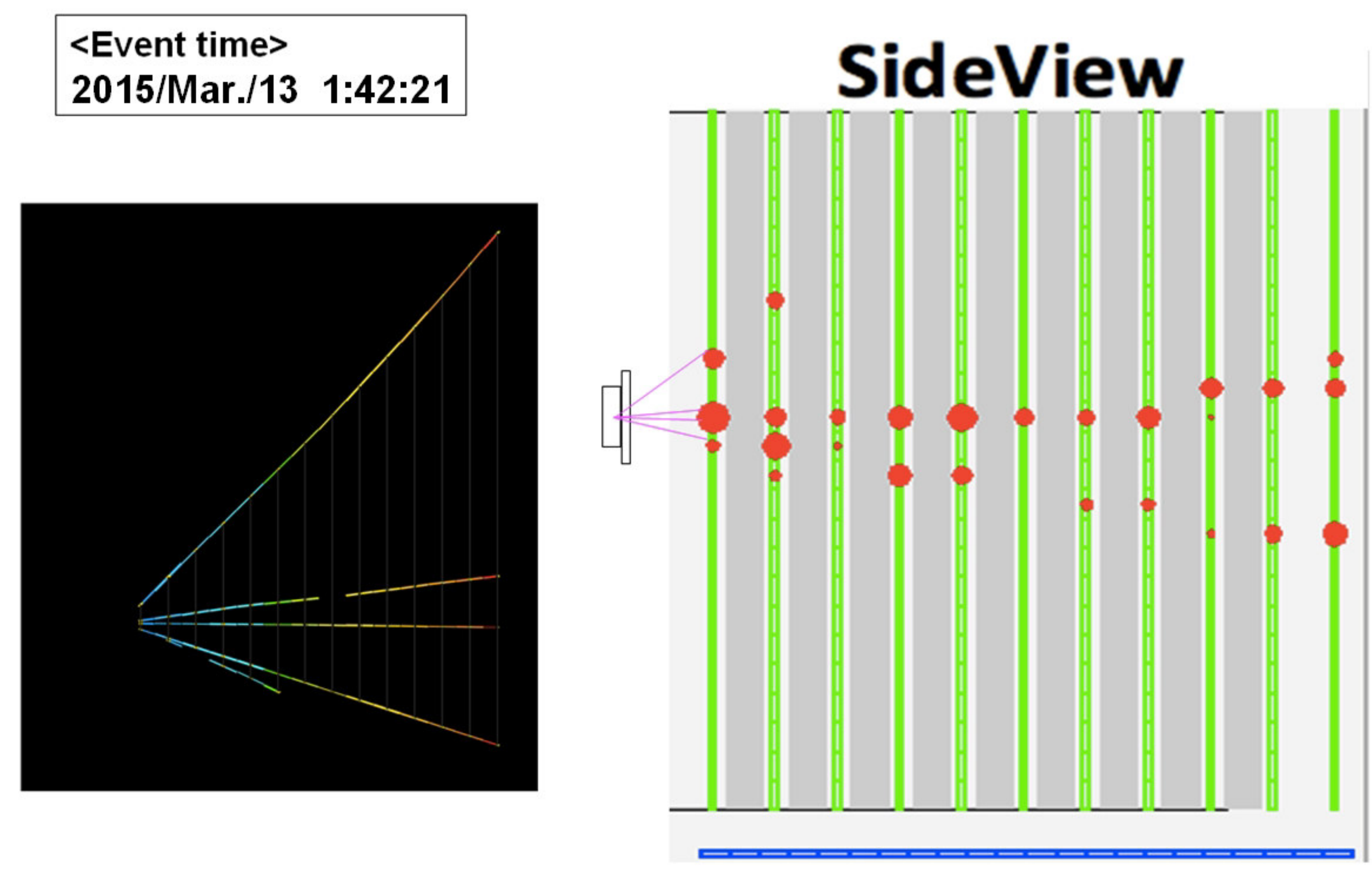}
\caption{Event.3: Hybrid analysis with INGRID}
\label{ev2side}
\end{center}
\end{figure}

The vertex is steel region at upstream of plate.14 and its vertex point is $(x,y) = (93105.6$ $\mu$m, $48182.9$ $\mu$m$)$ in the ECC coordinate system. There are three MIP tracks and one black track at the vertex. The details of the event tracks are listed in table \ref{vtxinf2}. This event is also matched between the emulsion detector and INGRID with the timing information, as shown in Fig.\ref{ev2top} and Fig.\ref{ev2side}. The measured track momenta are consistent with the measured range in INGRID. Unfortunately, since two tracks penetrate through INGRID, it is difficult to decide which is the muon track from the information of INGRID.

\begin{table}[!h]
\caption{Track information in Event.3}
\label{vtxinf2}
\scalebox{0.7}[1.3]
\centering
\begin{tabular}{|c|c|c|c|c|c|c|c|c|}
\hline
PL & segment id & ${tan\theta_x}$ & ${tan\theta_y}$ & \shortstack{IP \\ (${\mu}$ m)} & PH$_{AV.}$ & VPH$_{AV.}$ & \shortstack{p${\beta}$ \\ (GeV/c)} & PID \\ 
\hline
14 & 312320 & -0.1029 & -0.2115 & 4.8 & 26.5 & 62.6 & 2.21$^{+5.42}_{-0.83}$ & muon/pion \\
14 & 312317 & -0.1830 & -0.0077 & 4.9 & 27.1 & 63.2 & 1.09$^{+0.36}_{-0.25}$ & pion/muon \\
14 & 312315 &  0.2550 &  0.1005 &  0.4 & 31.1 & 150.6 & 0.27$^{+0.06}_{-0.06}$ & proton \\
14 & 297419 &  0.7655 &  0.6744 &  9.3 & 23.1 & 39.1 & 0.30$^{+0.06}_{-0.06}$ & MIP \\
13 & 328835 & -0.3979 & -0.2846 & 57.8 & 24.3 & 49.0 & -- & e$^\pm$ \\

\hline
\end{tabular}
\end{table}

\begin{figure}[ht]
\begin{center}
\includegraphics[clip, width=10.0cm]{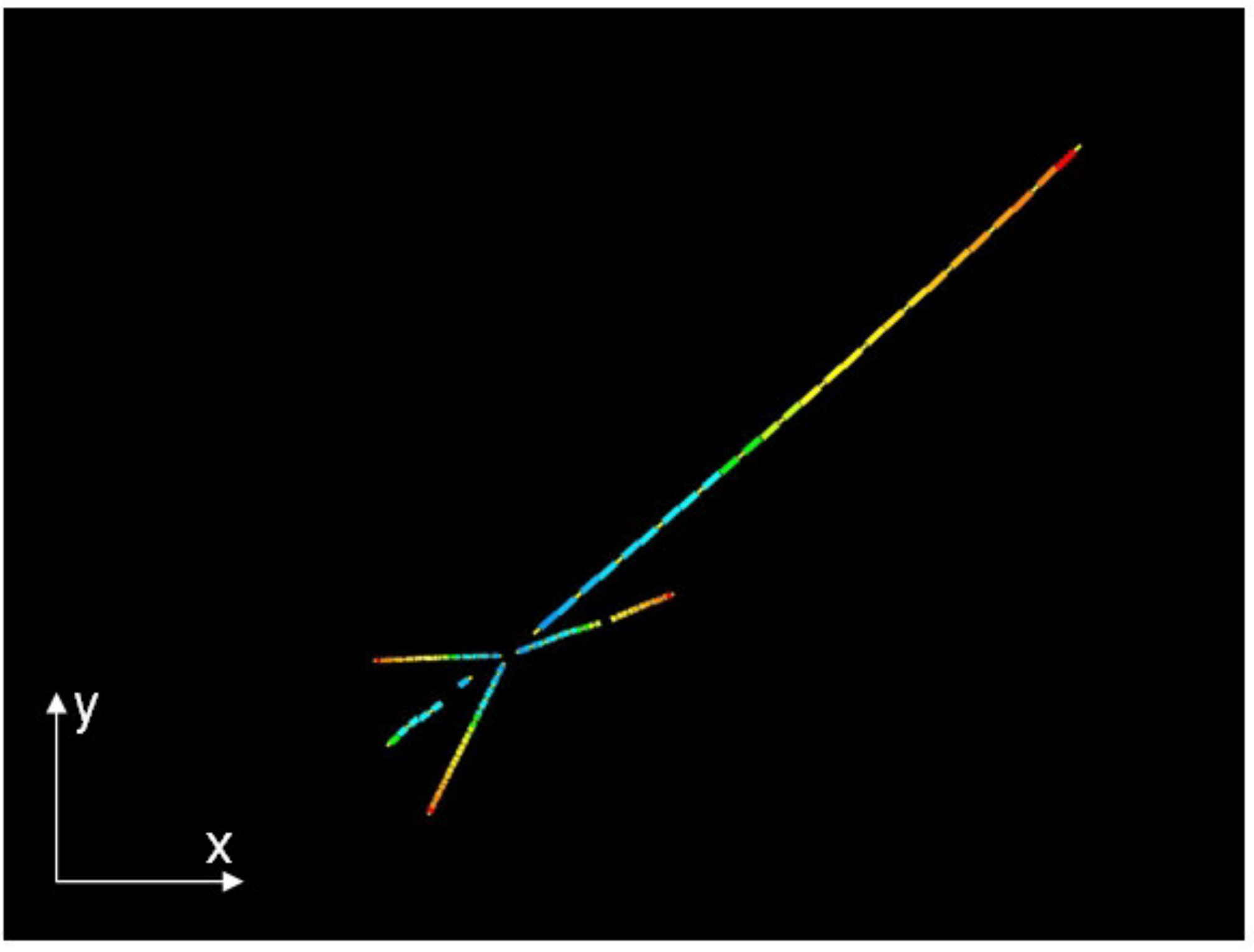}
\caption{Event.3}
\label{ev2z}
\end{center}
\end{figure}

\clearpage

\section{Future Prospects}

\subsection{Improvement of the Systematic Analysis}
Further neutrino events are going to be searched for in Categories (c), (d), and (h). It is important to reject many fake starting tracks to find new neutrino events. To do that, the energy threshold to connect tracks between emulsion films will be lowered. Another method is to reconstruct the electron/positron pairs of gamma-rays and neutron-interacted vertices from the backward direction. Then, the search for additional much larger angle black tracks will be also effective to identify the interaction point with large-angle track scanning \cite{LA}. Furthermore, when some tracks reached the emulsion shifter, it will be possible to take the coincidence with timing information for background rejection.

\subsection{Future Experimental Plan}
In 2016, anti-neutrino beam was exposed to 60 kg steel target ECC and film scanning was done. Currently the neutrino event analysis for this experiment is in progress to investigate the detector performance with higher statistics. Approximately 3,000 anti-neutrino interaction events are expected to accumulate in this ECC. Moreover, not only anti-muon neutrino events but also anti-electron neutrino events will be detected. Furthermore, the neutrino analysis will be proceeded in a few-kg water target emulsion chamber that is assembled with the vacuum-packed emulsion films and the frame-type spacers to fill the water as a target. The status and results of these analyses will be reported in the future. Then, a large-scale water target emulsion chamber will be constructed to study the neutrino--water interactions in detail.

\section{Conclusion}

We proposed a new experimental project with a nuclear emulsion detector to study low-energy neutrino--nucleus interactions in detail for future neutrino oscillation physics experiments. The J-PARC T60 experiment has been implemented as a first step of this program to check the feasibility and confirm the neutrino event analysis. In this work, the systematic neutrino event analysis with full ECC scanning data has been developed and the first neutrino event detection has been successfully demonstrated. Moreover, the hybrid analysis with INGRID has been performed for the first time. We will polish the systematic neutrino analysis chain, develop the water target emulsion chamber and steadily expand the detector scale to reach the desired physics output.

\section*{Acknowledgment}
We appreciate the support provided by the J-PARC Neutrino Group / Accelerator Group and the T2K Collaboration. We acknowledge the support from the Japan Society for the Promotion of Science (JSPS) through their grants (JSPS KAKENHI Grant Number 25105001, 25105006, 26105516, 26287049, 25707019, 20244031, 26800138, 16H00873).


%

\end{document}